\documentclass[physrev,superscriptaddress,
reprint
]{revtex4-2}
\usepackage[utf8]{inputenc}
\usepackage{upgreek}
\usepackage{graphicx}
\usepackage{dcolumn}
\usepackage{bm}
\usepackage[english]{babel}
\usepackage{amsmath}
\usepackage{amsfonts}
\usepackage{gensymb}
\usepackage[hidelinks]{hyperref}
\usepackage[all]{hypcap}
\usepackage{xcolor} 
\usepackage{soul} 
\usepackage{tabularx}
\usepackage{float}

\begin{document}
\author{Zoltán Kovács-Krausz}
\affiliation{Department of Physics, Institute of Physics, Budapest University of Technology and Economics, H-1111 Budapest, Hungary}
\affiliation{MTA-BME Superconducting Nanoelectronics Momentum Research Group, H-1111 Budapest, Hungary}
\author{Dániel Nagy}
\affiliation{Department of Physics of Complex Systems, ELTE Eötvös Loránd University, 1117 Budapest, Hungary}
\author{Albin Márffy}
\affiliation{Department of Physics, Institute of Physics, Budapest University of Technology and Economics, H-1111 Budapest, Hungary}
\affiliation{MTA-BME Superconducting Nanoelectronics Momentum Research Group, H-1111 Budapest, Hungary}
\author{Bogdan Karpiak}
\affiliation{Department of Microtechnology and Nanoscience, Chalmers University of Technology, SE-41296, Göteborg, Sweden}
\author{Zoltán Tajkov}
\affiliation{Hungarian Research Network, Centre for Energy Research, Institute of Technical Physics and Materials Science, 1121 Budapest, Hungary}
\affiliation{Centre of Low Temperature Physics, Institute of Experimental Physics, Slovak Academy of Sciences, Košice SK-04001, Slovakia}
\author{László Oroszlány}
\affiliation{Department of Physics of Complex Systems, ELTE Eötvös Loránd University, 1117 Budapest, Hungary}
\affiliation{MTA-BME Lendület Topology and Correlation Research Group, Budapest University of Technology and Economics, 1521 Budapest, Hungary}
\author{János Koltai}
\affiliation{Department of Biological Physics, ELTE Eötvös Loránd University, 1117 Budapest, Hungary}
\author{Péter Nemes-Incze}
\affiliation{Hungarian Research Network, Centre for Energy Research, Institute of Technical Physics and Materials Science, 1121 Budapest, Hungary}
\author{Saroj P. Dash}
\affiliation{Department of Microtechnology and Nanoscience, Chalmers University of Technology, SE-41296, Göteborg, Sweden}
\author{Péter Makk}
\email{makk.peter@ttk.bme.hu}
\affiliation{Department of Physics, Institute of Physics, Budapest University of Technology and Economics, H-1111 Budapest, Hungary}
\affiliation{MTA-BME Correlated van der Waals Structures Momentum Research Group, H-1111 Budapest, Hungary}
\author{Szabolcs Csonka}
\affiliation{Department of Physics, Institute of Physics, Budapest University of Technology and Economics, H-1111 Budapest, Hungary}
\affiliation{MTA-BME Superconducting Nanoelectronics Momentum Research Group, H-1111 Budapest, Hungary}
\author{Endre Tóvári}
\affiliation{Department of Physics, Institute of Physics, Budapest University of Technology and Economics, H-1111 Budapest, Hungary}
\affiliation{MTA-BME Correlated van der Waals Structures Momentum Research Group, H-1111 Budapest, Hungary}

\title{Signature of pressure-induced topological phase transition in ZrTe$_5$}

\begin{abstract}

The layered van der Waals material ZrTe$_5$ is known as a candidate topological insulator (TI), however its topological phase and the relation with other properties such as an apparent Dirac semimetallic state is still a subject of debate. We employ a semiclassical multicarrier transport (MCT) model to analyze the magnetotransport of ZrTe$_5$ nanodevices at hydrostatic pressures up to 2 GPa. The temperature dependence of the MCT results between 10 and 300\,K is assessed in the context of thermal activation, and we obtain the positions of conduction and valence band edges in the vicinity of the chemical potential. We find evidence of the closing and re-opening of the band gap with increasing pressure, which is consistent with a phase transition from weak to strong TI. This matches expectations from \textit{ab initio} band structure calculations, as well as previous observations that CVT-grown ZrTe$_5$ is a weak TI in ambient conditions.

\end{abstract}

\maketitle

\section{Introduction}

In certain materials, the spin-orbit interaction, through a process of band inversion, leads to a topological insulator (TI) phase\,\cite{Kane_QSHE_2005,topo-colloq}. The most remarkable consequence of this phase is that despite being gapped in the bulk, any boundary between a TI and a trivial (non-topological) insulator features gapless edge or surface states. These topological edge states are of interest in the area of quantum computation\,\cite{Sarma_2015_majorana}. In a two-dimensional (2D) case, this phase is also called a quantum spin Hall insulator. Initially, its existence had been proposed\,\cite{Bernevig_QSHE_2006,Bernevig_qshehgte_2006} and then successfully demonstrated in HgTe/CdTe quantum wells\,\cite{Konig_QSHEhgte_2007,Konig_QSHE_2008}. In addition, 3D TIs have also been discovered, initially with Bi$_2$Se$_3$ and Bi$_2$Te$_3$ at the forefront\,\cite{Xia_bi2te3_3dti_2009,Zhang_bi2te3_theory_2009}. These materials feature 2D surface states that have a chiral spin texture, where any orientation of the momentum uniquely determines the spin polarization, also called spin-momentum locking (SML)\,\cite{sml_and_supcond-topo-2011,sml-topo-2012,sml-topo-2014}. These chiral surface states are of interest for spintronics, since the SML mechanism can be taken advantage of, for example in heterostructures, to generate spin polarization and pass it into a spin transport medium such as graphene, or to electrically detect the presence of spin currents in the medium\,\cite{Vaklinova_grti_2016,optical_bi2se3_sml,khokhri-ree-2019}.

Zirconium pentatelluride (ZrTe$_5$) is a 2D layered material\,\cite{lattice_fjellvag-powDifExp-spzhpd_1986} and features an anomalous resistivity peak at a sample-dependent temperature $T_\text{p}$\,\cite{zrte5_resistanomaly_1981,zrte5_resistanomaly_1982}. This peak has been elucidated by angle-resolved photoemission spectroscopy (ARPES)\,\cite{bandstruct-and-fit_zhang-arpeslifshitz-eetilttnz_2017,zrte5tp_efshift_2018} to be caused by a shift of the chemical potential across the small band gap with temperature, while the predominant carriers change from holes to electrons as the temperature is reduced. In samples grown by chemical vapor transport (CVT), $T_\text{p}$ is typically in the $120-160$\,K range\,\cite{dirac-chiral_guolin-diracMagNeg-te3ddspz_2016,fit_shahi-bipolar-bcpoetp_2018,pressure_prb_akrap_2020}, allowing for study of both electron and hole transport properties based on temperature. The shift of the chemical potential and the values of $T_\text{p}$ have also been confirmed using \textit{ab initio} calculations when including doping\,\cite{zrte5_cvt_fourband}, which can be caused by Te vacancies as seen in CVT-grown samples\,\cite{fit_shahi-bipolar-bcpoetp_2018}.

ZrTe$_5$ is a candidate TI\,\cite{bandstruct_weng-QSHparadigm-tmpzh_2014}, with the monolayer expected to form a quantum spin Hall insulator. The multilayer structure could be in a weak TI (WTI) or strong TI (STI) state, depending on the $\mathbb{Z}_2$ invariant\,\cite{topo-colloq,sml_and_supcond-topo-2011}. The WTI case has surface states present only on certain boundaries, essentially acting like a stack of 2D TIs, while the STI case has topologically protected surface states on all boundaries, forming a genuinely 3D TI, as depicted in \hyperref[fig1]{Fig.~\ref{fig1}} (b) and (d), respectively. However, whether it forms a WTI or STI is under debate\,\cite{bandstruct_weng-QSHparadigm-tmpzh_2014,bandstruct_fan-WTIvSTI-tbswtizh_2017,dirac-and-sti-and-wti_mutch-strainTunedTransition-esttptz_2019,sdho-torus-flux_wang-magchiralaniso-gmatsz_2022}. Certain experimental evidence suggests an STI phase\,\cite{bandstruct_manzoni-STIarpes-estipz_2016,arpes_manzoni-ArpesSTI-tdnmbsz_2017,dirac-sti-and-sdho_wang-magnetoSdHO-mtestipz_2021}. However, most experiments have found CVT-grown ZrTe$_5$ to be a WTI with a smaller band gap\,\cite{bandstruct-and-fit_zhang-arpeslifshitz-eetilttnz_2017,dirac-and-sti-and-wti_mutch-strainTunedTransition-esttptz_2019,bandstruct-dirac_sun-pumpProbeDirac-pdsz_2020}, including observations using ARPES\,\cite{Zhang_ArpesWTI_2021} and scanning tunneling microscopy\,\cite{stmEdge_xbing-stmWTI-eoteds_2016,stmEdge_we-stmWTI-etesleg_2016}.

Theoretical band structure calculations generally predict a STI phase for ZrTe$_5$ and show the presence of several distinct band edges near the Fermi level\,\cite{bandstruct_weng-QSHparadigm-tmpzh_2014,bandstruct-and-fit_zhang-arpeslifshitz-eetilttnz_2017,bandstruct_fan-WTIvSTI-tbswtizh_2017}, supporting a multicarrier approach in the analysis of transport behavior. While the nonlinear Hall resistivity of ZrTe$_5$ has also been attributed to the anomalous Hall effect\,\cite{LiangAHEzrte5_2018,bandstruct-and-sdho-pressure_sun-AHEzeeman-lzsiahez_2020,gourgout_magneticAHEzrte5_2022}, recent gating experiments on thin flakes\,\cite{zrte5_thingating_2023} further support the multiband origin of the transport features. In addition, a few works consider the material a Dirac or Weyl semimetal instead\,\cite{dirac_chen-3Ddirac_msllzs3dmdf_2015,dirac_chen-3DdiracOptical_oss3ddsz_2015,dirac_xiang-diracMagOpt-oq2ddfz_2015,dirac-and-wti_chen-3DDiracIR-sebbi3dmd_2017}. The theoretical predictions and experimental observations generally agree that ZrTe$_5$ lies close to a WTI-STI phase transition boundary (yellow line in \hyperref[fig1]{Fig.~\ref{fig1}} (a)), with only a small band gap in the Dirac-like bands, potentially allowing for observation of relativistic behavior. In addition, the band structure is highly sensitive to small changes in lattice constants\,\cite{topo_tajkov-phase-strain_2022}. The goal of this work was to investigate whether this property could be exploited, by application of pressure, to drive the material through this phase transition.

Application of hydrostatic pressure is an emerging tool of interest in the study of 2D materials and heterostructures, with practical experimental methods currently in development\,\cite{Yankowitz_moirePressure_2018,pressure_fulop-vdW-pressure_2021}. The interlayer spacing is an important parameter, expected to tune properties of layered materials\,\cite{pressure_mos2_mi_2013,pressure_bilayer_2016} (including ZrTe$_5$\,\cite{bandstruct_weng-QSHparadigm-tmpzh_2014,bandstruct_fan-WTIvSTI-tbswtizh_2017}) and heterostructures\,\cite{TBLG_pressure_theory_2018,tajkov_uniaxial_bitexGraphene_2019}. Recently this tunability has been experimentally demonstrated in examples such as superconductivity in superlattices\,\cite{Yankowitz_moirePressure_2018}, band structure tuning of twisted bilayer graphene\,\cite{Yankowitz_tblgPressure_2019,Szentpeteri_tblg_pressure_2021}, enhancement of spin-orbit proximity effects\,\cite{balint_proximitySOI_2021,kedves2023stabilizing} and topological or magnetic ordering transitions\,\cite{pressure_cri3_afmfm,topo-magnet-pressure_yuan-eucd2as2_2022,topo-magnet-pressure_vohra-eusn2p2_2022}.

Although a few studies have been performed on bulk ZrTe$_5$ with application of pressure or strain\,\cite{pressure_pnas_mao_2016,sdho-nontrivial_zhang_dadss_2017,bandstruct-and-sdho-pressure_sun-AHEzeeman-lzsiahez_2020,pressure_prb_akrap_2020} (including regarding its phase transition\,\cite{dirac-and-sti-and-wti_mutch-strainTunedTransition-esttptz_2019}), experiments on thin nanodevices, where surface contributions are more evident in transport measurements, have not been performed before. The joint experimental and theoretical approach of this work is able to better reveal the signatures of this topological phase transition in ZrTe$_5$ nanoflakes.

\section{Theoretical Calculations and Multicarrier Model}

\begin{figure*}[!ht]
    \begin{center}
    \includegraphics[width=2.0\columnwidth]{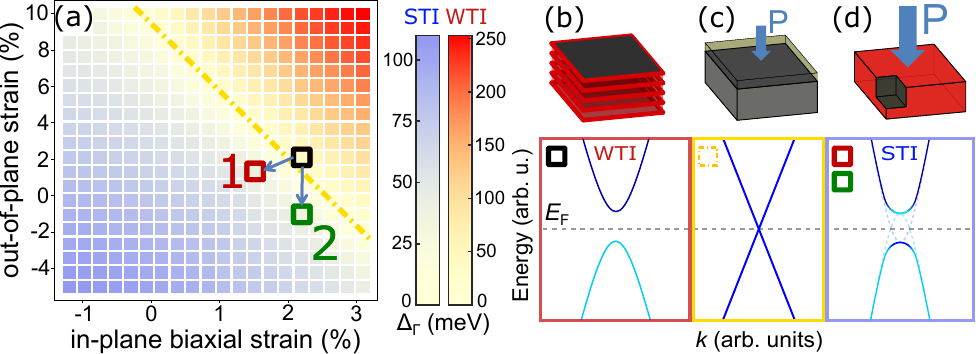}
    \caption{\textbf{Pressure-induced topological phase transition in ZrTe$_5$.}  (a) DFT-based strain map of ZrTe$_5$ showing the direct gap $\Updelta_\Upgamma$ as a function of lattice strain. The yellow dash-dotted line is the position where gapless Dirac bands are formed. The topological phase above and below this line is WTI and STI, respectively. Starting from the WTI phase in the black square, application of pressure (arrows) could lead to a WTI-to-STI phase transition, with two potential endpoints shown: an isotropic compression path (1, red square) and a compression in the vdW direction only (2, green square). (b) ZrTe$_5$ in a WTI phase corresponding to the black square in (a). A representative model of the bands along $\Upgamma$-Y is depicted below, showing a trivial gap. The individual ZrTe$_5$ layers are 2D TIs, forming distinct edge states (red contours) around the edges of the crystal. (c) At a certain pressure, the material reaches the transition point (yellow line as guide to the eye in (a)), and the gap closes. (d) Increasing the pressure, the STI phase is reached. The spin-orbit interaction leads to band inversion, and a gap opens up once again, forming a 3D TI that features surface states on each of the surfaces.}
    \label{fig1}
    \end{center}
\end{figure*}

First, in order to assess the topological phase of ZrTe$_5$ as a function of changes to lattice constants, and to help interpret the experimental results, we have performed \textit{ab initio} band structure calculations, employing density functional theory (DFT). For the detailed methodology, consult Supplementary Note 8. The size of the direct gap at the $\Upgamma$ point, $\Updelta_\Upgamma$, is plotted as a function of in-plane and out-of-plane strain in \hyperref[fig1]{Fig.~\ref{fig1}} (a). The phase transition, where $\Updelta_\Upgamma=0$, is found along the yellow dash-dotted line. The topological phase of the system is found to be in a WTI phase above this line and an STI phase below it. The relaxed lattice constants are $a$\,=\,2.002\,\AA, $b$\,=\,7.204\,\AA\,and $c$\,=\,13.876\,\AA\,\cite{topo_tajkov-phase-strain_2022}, representing the zero strain position on the strain map. These parameters fall within 1.1\% of those obtained from X-ray diffraction\,\cite{lattice_fjellvag-powDifExp-spzhpd_1986}. This discrepancy may be related to the presence of Te vacancies in CVT-grown ZrTe$_5$\,\cite{fit_shahi-bipolar-bcpoetp_2018}. It is important to note that the DFT-based band structure obtained using the relaxed constants does not accurately represent the experimental ZrTe$_5$ results on CVT-grown crystals, instead predicting a larger $\Updelta_\Upgamma$ (around 80 meV), and, most notably, resulting in a STI rather than WTI phase. According to Ref.\,\cite{zrte5_multicarrier_kkz}, the experimentally estimated ZrTe$_5$ bands are better represented by the DFT results at the black square in \hyperref[fig1]{Fig.~\ref{fig1}} (a), which is close to the phase transition boundary and consistent with the experimental results of Refs.\,\cite{bandstruct-and-fit_zhang-arpeslifshitz-eetilttnz_2017,dirac-and-sti-and-wti_mutch-strainTunedTransition-esttptz_2019,bandstruct-dirac_sun-pumpProbeDirac-pdsz_2020,Zhang_ArpesWTI_2021,stmEdge_xbing-stmWTI-eoteds_2016,stmEdge_we-stmWTI-etesleg_2016} suggesting a WTI phase.

A visual representation of the vicinity of the $\Upgamma$ point for the WTI phase is shown on the lower part of \hyperref[fig1]{Fig.~\ref{fig1}} (b), having a trivial gap without band inversion for the 3D band structure. In this state, the edge states of the WTI phase run along the edges of the individual ZrTe$_5$ layers, essentially each of them forming a 2D topological insulator, as is shown in the upper part of panel (b). Because this position is close to the gapless phase transition point, application of pressure is expected to compress the lattice and move the system through the phase transition and into an STI phase. The exact compression path is hard to predict, but the expectation from hydrostatic compression would be a movement from upper right to lower left of \hyperref[fig1]{Fig.~\ref{fig1}} (a) (arrows). For visualization, two example paths are chosen, labeled as 1 (2) showing an isotropic (vdW-only) compression process leading to an end position depicted as the red (green) square, respectively. Under the compression process, at a particular pressure value $p_\text{t}$, the material would reach the transition point depicted in \hyperref[fig1]{Fig.~\ref{fig1}} (c), where the gap between the Dirac-like bands closes. Note that the overall band structure often has zero indirect gap, even away from the phase transition point, due to the multiple side bands being close to the chemical potential. For this reason, it is important to be able to assess the direct gap $\Updelta_\Upgamma$ related to the topological phase transition. Further increasing the pressure would move the material into the STI phase, represented in \hyperref[fig1]{Fig.~\ref{fig1}} (d). Here, a band gap opens up again, but in this case the spin-orbit interaction leads to band inversion, highlighted by the coloring near the band edges. Gapless surface states will be found crossing this gap at every boundary (surface) of the crystal. Therefore, an experimental observation of $\Updelta_\Upgamma$ closing and reopening can be considered as a signature of the topological phase transition.

For experimental investigation under pressure, nanoflakes of ZrTe$_5$ have been obtained by mechanical exfoliation to a SiO$_2$ substrate. Predominantly, flakes in the $50-150$ nm thickness range have been studied, but measurements have also been performed on flakes thinner than 50 nm to study thickness dependence. The ZrTe$_5$ crystals favorably cleave into regular rectangular single crystals of several $\upmu$m dimensions, where the longer edge of the exfoliated crystals is parallel to the crystallographic $a$-axis\,\cite{dirac_xiang-diracMagOpt-oq2ddfz_2015,multiband-and-sdho_gang_ooeipahmflz_2016}. This allowed for easy establishment of Hall-bar measurement geometries on the flakes by depositing Cr/Au metallic contacts. A novel method for measuring nanodevices-on-chip under hydrostatic pressure was employed\,\cite{pressure_fulop-vdW-pressure_2021}, in a piston-cylinder pressure cell that fits into the variable temperature insert of a liquid helium cyrostat with superconducting magnet (with fields up to 8\,T), allowing for a full range ($1.5 - 300$\,K) of temperature-dependent magnetotransport measurements at pressures up to 2 GPa.

\begin{figure*}[!ht]
    \begin{center}
    \includegraphics[width=2.0\columnwidth]{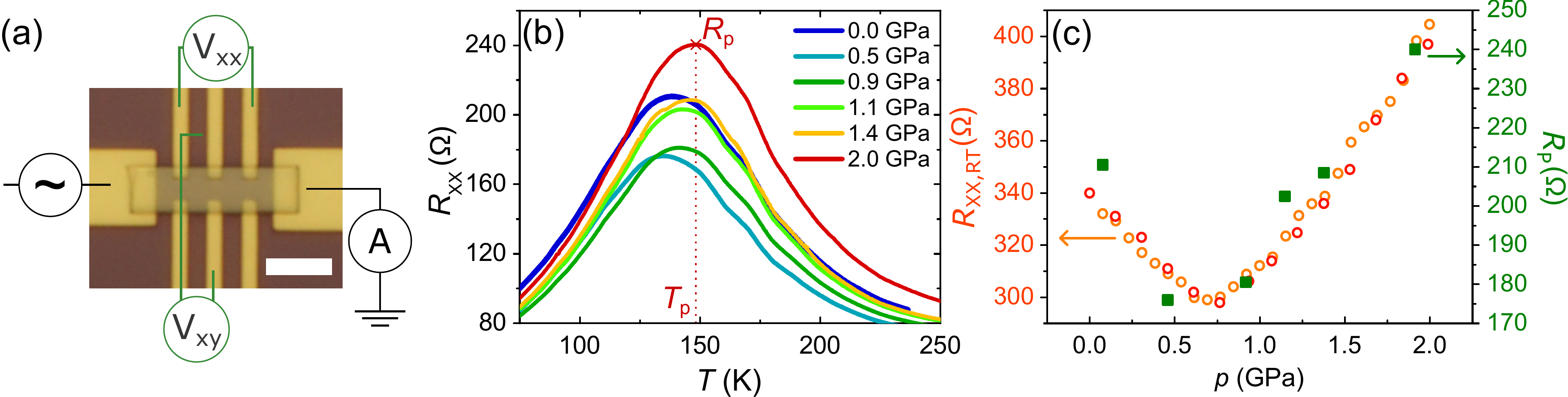}
    \caption{\textbf{Transport measurements on ZrTe$_5$ under pressure.}  (a) Optical image of a ZrTe$_5$ nanodevice (sample S2, 101 nm thick) on SiO$_2$ substrate with metallic contacts. The scale bar is 5 $\upmu$m. The Hall-bar-like measurement setup enables simultaneous measurement of $R_\text{xx}$ and $R_\text{xy}$ for magnetotransport. (b) Temperature dependence of $R_\text{xx}$ of sample S1 at various pressures, showing a non-monotonic dependence of the resistivity peak magnitude $R_\text{p}$ and its position $T_\text{p}$. (c) Sample resistance measured in a different sample (S6) at room temperature (RT) during dynamic pressure application and release (orange and red circles). The pressure dependence of $R_\text{p}$ extracted from (b), shown as green squares, correlates with that of $R_\text{xx}$ measured at RT.}
    \label{fig2}
    \end{center}
\end{figure*}

A multicarrier transport (MCT) approach was used to analyze the magnetotransport measurements and obtain band structure details under pressure. We have obtained the sheet conductivity data $\sigma_\text{xx}$ and $\sigma_\text{xy}$ from longitudinal and Hall resistance $R_\text{xx}$ and $R_\text{xy}$ measured in an out-of-plane $B_\text{z}$ magnetic field. Following Ref.\,\cite{zrte5_multicarrier_kkz}, the obtained sheet conductivity data was fitted using the MCT model, using signed conductivity $\sigma_\text{i}$ and mobility $\mu_\text{i}$ as the fitting parameters for the \textit{i}-th carrier. The procedure was performed at each temperature by fitting the following equations simultaneously using the same $\sigma_\text{i}$ and $\mu_\text{i}$ parameters:

\noindent\begin{minipage}{1.0\linewidth}
\begin{align}
		\sigma_\text{xy}(B)  = \sum_{\text{i=1}}^{\text{NC}} \frac{\sigma_\text{i}\mu_\text{i}B}{1+\mu_\text{i}^{2}B^{2}}, \, \sigma_\text{xx}(B)  = \sum_{\text{i=1}}^{\text{NC}} \frac{\lvert \sigma_\text{i} \rvert}{1+\mu_\text{i}^{2}B^{2}}.
	\label{mctsigma1}
	\end{align}
\end{minipage}

Each fit is performed several times using randomized starting parameters, to ensure the robustness of the results. The required number of carriers (NC) to adequately fit the conductivity curves was found to vary with temperature between $3-5$, with the highest number being necessary near the anomalous resistivity peak $T_\text{p}$, where the chemical potential is crossing the small band gap. The fitting is demonstrated at several temperatures in Supplementary Note 2. The electron or hole nature of a carrier is fixed throughout the temperature range, i.e. $\sigma_\text{i}$ cannot change sign. Thus, the obtained fitting results exhibit a realistic freeze-out behavior expected from the shifting chemical potential in ZrTe$_5$, as will be demonstrated below in relation to \hyperref[fig4]{Fig.~\ref{fig4}} (c) and (e). As shown in Ref.\,\cite{zrte5_multicarrier_kkz}, one of the carriers (by convention $i=1$) is an edge-confined carrier. This can be envisioned as a diffusive 2D surface state running only along the vertical sides of the ZrTe$_5$ crystal. This surface state is parallel to the applied $B_\text{z}$ and is thus unaffected by it. In the MCT model this is implemented by setting its $\sigma_1$ to a finite value while having $\mu_1=0$, resulting in a constant, field-independent contribution to $\sigma_\text{xx}$, and zero $\sigma_\text{xy}$. This edge carrier is necessary to obtain smooth and realistic temperature trends for the densities and mobilities of all carriers. For all carriers other than this edge carrier, the $\sigma_\text{i}=n_\text{i}q_\text{i}\mu_\text{i}$ relation applies, where $n_\text{i}$ is the 2D density of the $i$-th carrier and $q_\text{i} = \pm e$ is its charge. Dividing the 2D carrier density $n_\text{i}$ with crystal thickness results in the 3D carrier density $n_\text{3D,i}$.

Since in ZrTe$_5$ the chemical potential crosses the band gap with temperature, the obtained $n_\text{3D,i}(T)$ data of the individual carriers can be used to fit a model band edge corresponding to the carrier, which has the form:

\begin{align}
		n_\text{3D,i}(T)=\int_{\varepsilon}g(\varepsilon,E_\text{i},m^*_\text{i})f(\varepsilon,T,r_\text{i})d\varepsilon,
	\label{ntintegral1}
\end{align}
where $g$ is the density of states (DOS) of a band having the band edge energy $E_\text{i}$ and effective mass $m^*_\text{i}$, assuming a 3D isotropic and quadratic band near the band edge. $f$ is the Fermi function with a temperature-dependent chemical potential, which is shifted with a rate $r_\text{i}=0.25$\,meV\,K$^{-1}$ with temperature (from Ref.\,\cite{bandstruct-and-fit_zhang-arpeslifshitz-eetilttnz_2017}). We note that, similarly to Ref.\,\cite{zrte5_multicarrier_kkz}, for the two high-mobility carriers (which we attribute to the ZrTe$_5$ bands near $\Upgamma$) we have used a 3D Dirac DOS for the fitting, with the Fermi velocity $v_\text{F,i}$ taking the place of $m^*_\text{i}$. Most importantly, this method allows for the extraction of the direct gap at $\Upgamma$ from the magnetotransport data, by subtracting the obtained band edge energies of the conduction and valence bands represented by these two carriers.

\begin{figure*}[!ht]
    \begin{center}
    \includegraphics[width=2.0\columnwidth]{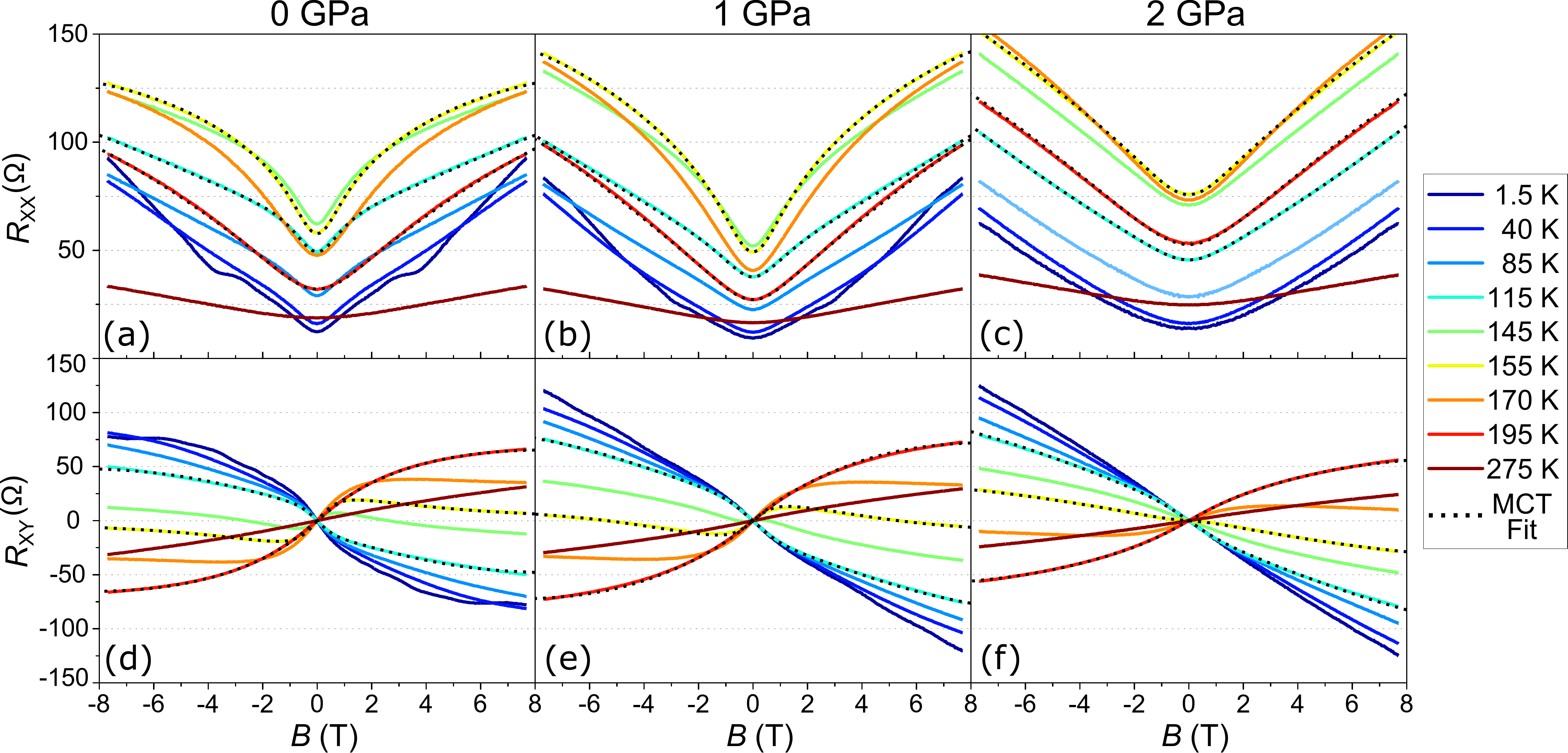}
    \caption{\textbf{Magnetotransport measurements on ZrTe$_5$ under pressure.} (a)-(c) $R_\text{xx}$ trends at select temperatures on both sides of $T_\text{p}$ (approximately 152 K) for sample S2 at approximately 0, 1 and 2 GPa pressures, respectively (pressure increases from left to right). (d)-(f) $R_\text{xy}$ trends of the same device at these pressures. For temperatures $T=115$, $155$ and $195$\,K, the resistances corresponding to the fitting results of the MCT model are overlayed on the experimental curves as black dotted lines. }
    \label{fig3}
    \end{center}
\end{figure*}

\section{Experimental Results and Discussion}

Transport measurements on ZrTe$_5$ nanoflakes, including application of pressure, were successfully performed on six devices in the $50-150$ nm thickness range, all of which behaved consistently. Here we discuss the results on two example devices; data on the others is shown in Supplementary Note 1. The measurement setup is shown in \hyperref[fig2]{Fig.~\ref{fig2}} (a), where the ZrTe$_5$ crystallographic $a$-axis is oriented horizontally on the image. A bias current was applied and simultaneous longitudinal and transverse measurements were taken. The temperature dependence of $R_\text{xx}$, the longitudinal resistance of device S1 (approx. 155\,nm thick), is plotted at different pressures in \hyperref[fig2]{Fig.~\ref{fig2}} (b). The characteristic resistivity peak of ZrTe$_5$ is observed to change with pressure, with both the resistance magnitude $R_\text{p}$ and its temperature position $T_\text{p}$ changing non-monotonically, first showing a decrease and then a subsequent increase with application of pressure. In \hyperref[fig2]{Fig.~\ref{fig2}} (c) the extracted $R_\text{p}$ pressure dependence is shown as green squares. Though temperature dependent measurements can only be performed at a static pressure, which is set at room temperature (RT), the RT sample resistance can be measured dynamically during pressure application or release. The RT pressure dependence of sample S6 (see magnetotransport data in Supplementary Fig. 1) is plotted as the orange datapoints in \hyperref[fig2]{Fig.~\ref{fig2}} (c), and follows the $R_\text{p}$ pressure dependence trend of sample S1 (green squares) well. This RT pressure dependence of $R_\text{xx}$ is robust and reproducible across all samples, with a distinguishable minimum around 0.7 GPa.

Our observation of a minimum in $R_\text{p}$ with increasing pressure is consistent with that of the bulk CVT-grown sample in Ref.\,\cite{pressure_prb_akrap_2020} (see their Fig. 2 (c)). In comparison, Ref.\,\cite{sdho-nontrivial_zhang_dadss_2017} on a similarly grown bulk sample does not feature a $R_\text{p}$ minimum, although the first shown non-zero pressure is 0.8 GPa, above our observed $R_\text{xx}$ minimum at 0.7 GPa. Meanwhile, Ref.\,\cite{bandstruct-and-sdho-pressure_sun-AHEzeeman-lzsiahez_2020} measured flux-grown ZrTe$_5$ under pressure, with a $T_\text{p}$ close to 2\,K, making the results not directly comparable. Higher pressure studies were performed in Ref.\,\cite{pressure_pnas_mao_2016} using a diamond anvil cell. Their results in the $0-2$\,GPa pressure range (see their Fig. 1 (a)) show an overall reduction of the peak resistance $R_\text{p}$, in contrast with our measurements and those of Refs.\,\cite{pressure_prb_akrap_2020,sdho-nontrivial_zhang_dadss_2017} which show an overall increase.

In order to employ the MCT method and obtain band structure details under pressure, magnetotransport measurements were performed. \hyperref[fig3]{Fig.~\ref{fig3}} shows a subset of the measurements in the $1.5 - 300$\,K range on device S2 (approx. 98\,nm thick) at pressures of approximately 0, 1 and 2\,GPa respectively. Since the pressure medium freezes in the studied temperature range, the pressure decreases by roughly 0.1\,GPa between 300 and 1.5\,K when having a RT starting pressure above 1\,GPa\,\cite{daphne_pressuredep}. Therefore we consider 0.1\,GPa as the uncertainty value. The $R_\text{xx}$ and $R_\text{xy}$ data have been symmetrized and anti-symmetrized, respectively. The 0\,GPa dataset features the typical complex magnetotransport behavior of CVT-grown ZrTe$_5$\,\cite{bandstruct-and-fit_liu-dynmass-zsdmgdsz_2016,multiband-and-fit_lu-thicknessDep-tttbtns_2017,fit_shahi-bipolar-bcpoetp_2018,zrte5_multicarrier_kkz}, with S-like shapes in $R_\text{xy}$ originating from the the multiple carriers involved in transport. The gradual transition from valence to conduction band with decreasing temperature can be seen in the sign reversal of the predominant slope of $R_\text{xy}$. There is a noticeable effect with increasing pressure, which can be described as an apparent "flattening" and simplification of both $R_\text{xx}$ and $R_\text{xy}$ curves at most temperatures. For example, the yellow $R_\text{xy}$ curves (close to $T_\text{p}$) in \hyperref[fig3]{Fig.~\ref{fig3}} (d)-(f) initially have a slope reversal at low field, but this vanishes at 2 GPa and the slope is negative throughout. A notable change in the low temperature (blue) $R_\text{xy}$ curves is that their hyperbolic-tangent-like shape at 0 GPa becomes almost completely linear at a pressure of 2 GPa. In addition, the weakening of Shubnikov-de Haas oscillations with increasing pressure can be observed in $R_\text{xx}$ in \hyperref[fig3]{Fig.~\ref{fig3}} (a)-(c) (dark blue $T=1.5$\,K curves). Overall, the $R_\text{xx}$ curves at most temperatures, but especially near $T_\text{p}$, feature a prominent symmetric dip at low pressure (panel (a)), and become closer to a parabolic- or V-shape under high pressure (panel (c)).

\begin{figure*}[!ht]
    \begin{center}
    \includegraphics[width=2.0\columnwidth]{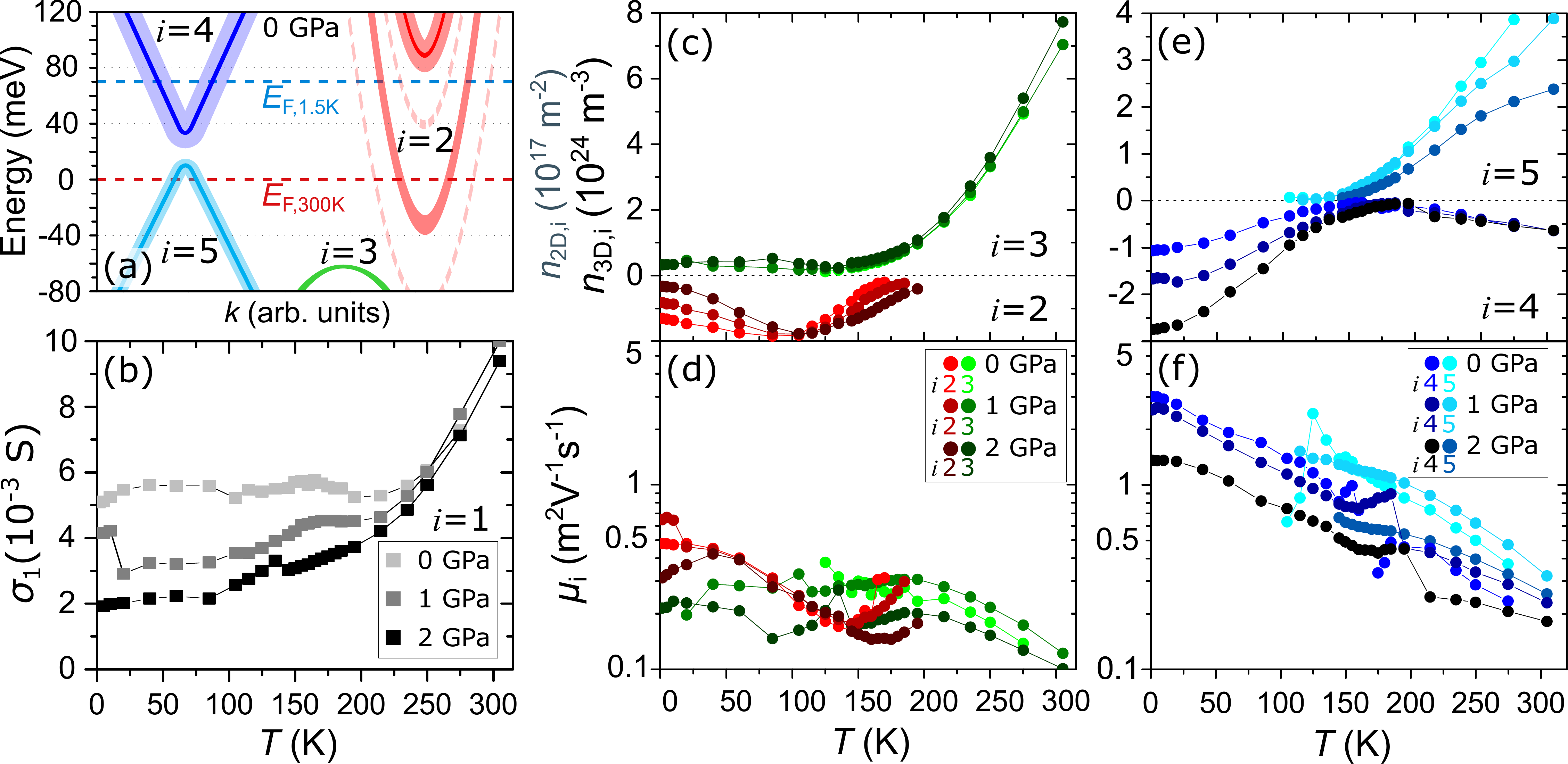}
    \caption{\textbf{Multicarrier fitting results.}  (a) Visual representation of the band fitting results of the 0\,GPa dataset of sample S2. The k-axis is not to scale, only the band edge energies are used. Bands $i=2,3$ are parabolic while $i=4,5$ are Dirac bands representing the higher mobility carrier pair. The size of the direct gap $\Updelta_\Upgamma$ can be extracted from the latter. (b) The $\sigma_1$ contribution of the edge carrier ($i=1$) decreases with pressure. (c),(d) Carrier density $n_\text{3D,i}$ (signed) and mobility $\mu_\text{i}$ data, respectively, of the two low-mobility carriers ($i=2,3$). The y-axis of (c) also gives the scale of the corresponding sheet density $n_\text{2D,i}$, which is $n_\text{3D,i}$ multiplied by the device thickness. The evolution with pressure is shown, where higher pressures are represented by darker colors. (e),(f) The same data for the two high-mobility Dirac-like carriers ($i=4,5$).}
    \label{fig4}
    \end{center}
\end{figure*}

\begin{figure*}[!ht]
    \begin{center}
    \includegraphics[width=1.9\columnwidth]{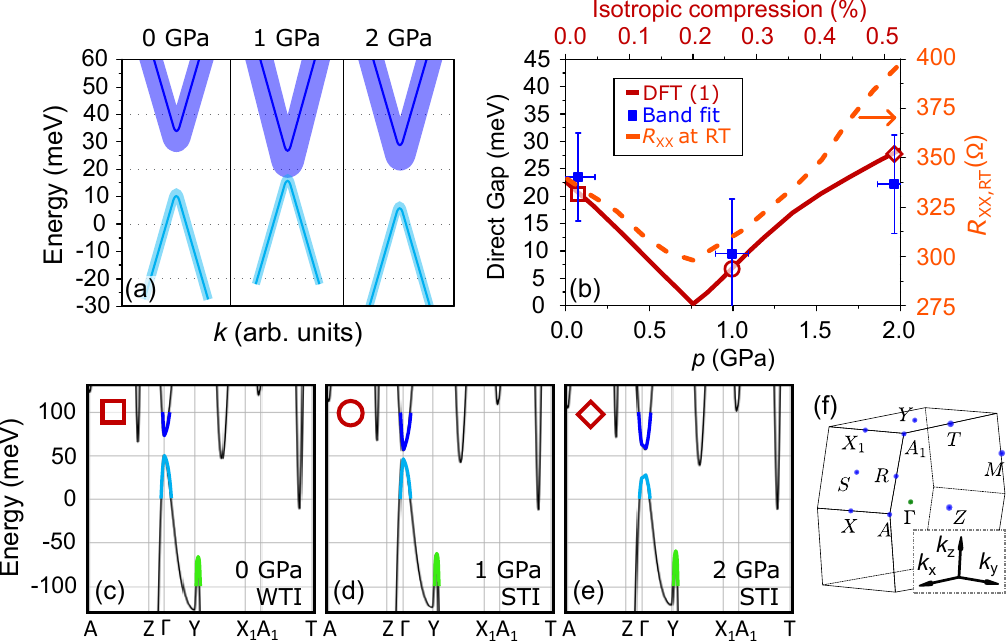}
    \caption{\textbf{Band fitting results and evolution of direct gap $\Updelta_\Upgamma$ with pressure.} (a) The position of the band edges at 300\,K, obtained from the carrier fitting results on sample S2. The error is represented by the thickness of the respective lines in lighter color. $\Updelta_\Upgamma$ can be extracted as the band edge difference of the two high-mobility (blue) bands. (b) The plotted value and error of $\Updelta_\Upgamma$ with pressure (blue squares). The dynamic response of the sample resistance with pressure, measured at RT, is plotted as an orange dashed line. The red solid line represents the evolution of $\Updelta_\Upgamma$, extracted from DFT calculations, along an isotropic compression path (upper x-axis). It has been scaled to the (lower) pressure axis by matching the curve minimum ($\Updelta_\Upgamma=0$) to the minimum of the $R_\text{xx}$ measurement. (c)-(e) Calculated band structures corresponding to the 0, 1 and 2 GPa pressures, respectively, along the isotropic compression path (red DFT (1) curve in (b)). Their positions on the path are marked in (b) as a red square, circle and diamond, respectively. The green band edges correspond to the $i=3$ carrier of the MCT model. (f) Brillouin Zone of ZrTe$_5$, with high-symmetry points that have been used for the x-axis of (c)-(e).}
    \label{fig5}
    \end{center}
\end{figure*}

Following the methodology outlined in Sec. II, the magnetotransport curves from \hyperref[fig3]{Fig.~\ref{fig3}} can be fitted with the MCT model (\hyperref[mctsigma1]{Eqs.~\ref{mctsigma1}}). The effects of pressure on the MCT fitting are demonstrated in detail in Supplementary Note 3. For a few selected temperatures, the fitted curve (the $\sigma_\text{xx(xy)}$ of the fit result) has been transformed back into $R_\text{xx(xy)}$ and plotted as black dotted lines in \hyperref[fig3]{Fig.~\ref{fig3}}. The results of the fitting along the full temperature range are shown in \hyperref[fig4]{Fig.~\ref{fig4}}. The signed carrier density ($n_\text{3D,i}$) trends with temperature are plotted for the low-mobility ($i=2,3$, panel (c)) and high-mobility ($i=4,5$, panel (e)) carriers. The corresponding mobilities ($\mu_\text{i}$) are plotted below, in \hyperref[fig4]{Fig.~\ref{fig4}} (d) and (f), respectively. It can be seen that each panel features an electron type ($i=2,4$, with $n_\text{3D}<0$) and a hole type ($i=3,5$, with $n_\text{3D}>0$) carrier, with the dominant carrier type changing near $T_\text{p}$, in a manner consistent with the expectation of freeze-out due to the shift of the chemical potential with temperature. This is most easily visible in panel (e) corresponding to the high-mobility Dirac-like bands near $\Upgamma$. A notable pressure dependence can be seen in this high-mobility pair. For the hole type carrier ($i=5$), visible at higher temperatures, we observe a decrease of its $n_\text{3D}$ amplitude with pressure. Meanwhile, the electron carrier ($i=4$), which becomes relevant at lower temperatures, shows the opposite, an increase in the density with pressure. Note that at the lowest temperatures, where Landau level physics and phase coherence effects become relevant, the results of the MCT fitting should not be considered accurate. The $\sigma_1$ contribution of edge-confined carrier is shown in \hyperref[fig4]{Fig.~\ref{fig4}} (b), exhibiting a notable decrease with increasing pressure. This carrier has no associated mobility value.

Fitting of the $n_\text{3D,i}$ data using \hyperref[ntintegral1]{Eq.~\ref{ntintegral1}} results in the band edge energies $E_\text{i}$, where most importantly the difference of the two high-mobility bands ($i=4,5$) can be used to get an estimate for $\Updelta_\Upgamma$. The final extracted band edge energies of the obtained bands at 0 GPa, relative to the chemical potential at 300\,K, are represented in \hyperref[fig4]{Fig.~\ref{fig4}} (a). The darker, thin colored lines show the band edges attributed to each carrier, and the thicker, lighter colors represent the error of the fit. Note that for the $i=2$ carrier there are two solid bands drawn, because, according to the fitting procedure, this band shifts faster with temperature than the others. Therefore, the upper red band shows its starting position at 300\,K while the lower band shows its final position at low temperature. This result is similar to the ARPES results of Ref.\,\cite{bandstruct-and-fit_zhang-arpeslifshitz-eetilttnz_2017}, showing an electron pocket along the A$_1$-T high symmetry line shifting at a higher rate with temperature. The dashed red bands represent the error of the shift rate, marking the minima and maxima of the final band position at 1.5\,K. The other bands in our results shift at a rate of 0.25\,meV\,K$^{-1}$ (according to Ref.\,\cite{bandstruct-and-fit_zhang-arpeslifshitz-eetilttnz_2017}), which is already represented by the $E_\text{F}$ shift in \hyperref[fig4]{Fig.~\ref{fig4}} (a). Apart from an offset (which could be explained by n-doping), the obtained band edges are in good agreement with the DFT-based band structure from the black square position in \hyperref[fig1]{Fig.~\ref{fig1}} (a), see also \hyperref[fig5]{Fig.~\ref{fig5}} (c). 

The pressure dependence of the gapped structure at $\Upgamma$ is represented in \hyperref[fig5]{Fig.~\ref{fig5}} (a), showing the band edges attributed to the $i=4$ and $i=5$ carriers. The extracted gap size $\Updelta_\Upgamma$, and its error, are shown in the plot of \hyperref[fig5]{Fig.~\ref{fig5}} (b) as blue squares. The pressure evolution of the gap shows a decrease and subsequent increase, as would be expected from passing a WTI-to-STI phase transition. The tendency of the resistivity with pressure (measured at RT) is shown as the orange dashed line on the same plot, with a notable minimum. We suspect that this minimum, along with a similar minimum in $R_\text{p}$ and overall temperature-dependent resistivity (see \hyperref[fig2]{Fig.~\ref{fig2}}), is consistent with the topological phase transition. In Ref.\,\cite{dirac-and-sti-and-wti_mutch-strainTunedTransition-esttptz_2019} a similar resistivity minimum was observed while applying uniaxial strain to bulk ZrTe$_5$ and is attributed to a WTI-STI transition. The gapless state would therefore correspond to the position of this minimum, while opening a gap on either side of the phase transition would lead to an increase in resistivity. This is expected even more so at temperatures above $T_\text{p}$, where the chemical potential is still in the valence band, and transport contributions due to thermal excitations across the gap are relevant. This would explain the observed non-monotonic pressure dependence. To verify the evolution of the gap size with pressure using the DFT-based calculations, we have extracted the gap size along the isotropic pressure path from \hyperref[fig1]{Fig.~\ref{fig1}} (a), plotted in red. The x-axis of the DFT-based compression path (note the upper x-axis of \hyperref[fig5]{Fig.~\ref{fig5}} (b)) has been scaled to the pressure axis such that the gapless position is at the resistivity minimum of the devices. It can be observed that the DFT-based gap size matches the experimentally obtained data reasonably well on both sides of the expected phase transition. A comparison of the isotropic (1) and vdW-only (2) compression paths from \hyperref[fig1]{Fig.~\ref{fig1}} (a) can be found in Supplementary Note 7. The result suggests that the exact choice of compression path makes little difference.

The DFT-based band structures corresponding to the 0, 1 and 2 GPa pressures are shown in \hyperref[fig5]{Fig.~\ref{fig5}} (c)-(e), while the Brillouin zone of ZrTe$_5$ and the relevant high-symmetry points are shown in panel (f). The band structure starts in WTI phase at 0 GPa but is already in STI phase at 1 GPa. From the RT resistivity dependence, the transition is expected at $p_\text{t} \approx 0.7$ GPa corresponding to an isotropic compression of around 0.2\%. The gap opens further with the increase to 2 GPa, where the shape of the inverted bands at $\Upgamma$ is better visible. Importantly, the band edges defining the gap at 2 GPa are at a lower energy than at 0 GPa, which can also be seen in the experimental band edge results of \hyperref[fig5]{Fig.~\ref{fig5}} (a): despite the 0 and 2 GPa gaps being of approximately similar size, both band edges have moved to a lower position. Finally, one may note the presence of electron sidebands along the A1-T line, (to which we attribute the $i=2$ carrier in the MCT model), which extend to lower energies than the valence band edge, meaning that the entire band structure is gapless.

Finally, we have also investigated magnetotransport in devices thinner than 50 nm, in the range where transport properties start to vary significantly with thickness. Here we shortly discuss our conclusions, while the detailed results are presented in Supplementary Note 4. In general, the thinner samples exhibit hole-dominated transport throughout the temperature range, as noted before in existing literature\,\cite{multiband-and-sdho_gang_ooeipahmflz_2016,multiband-and-fit_lu-thicknessDep-tttbtns_2017,multiband-nano_jj-etnzs_2017,thicknessDepChn_tang-mctzf_2018}. This suggests the chemical potential is shifted well inside the valence band at 300\,K, hindering the transition to electron transport as temperature decreases. The MCT model and band fitting confirms this, as the band edge of the hole type carriers is situated at higher energies for thinner samples, with a monotonic thickness dependence (see Supplementary Note 5). We have successfully investigated the pressure response of a 22 nm thick crystal (presented in Supplementary Note 6). The minimum in its $R_\text{xx}(p)$ dependence at RT is still clearly visible, at a slightly higher pressure than seen in thicker devices, closer to 1 GPa. Therefore, it is expected that crystals of this thickness would still undergo a phase transition, although the fitting of \hyperref[ntintegral1]{Eq.~\ref{ntintegral1}} is no longer viable and cannot reveal the gap size due to the increased distance of the chemical potential from the band edges. Improved gating techniques, such as ionic liquid gating, employed on thin ZrTe$_5$, could be an avenue of further investigation. Shifting the chemical potential and restoring the electron-hole carrier change behavior with temperature to such devices would allow for better assessment of the thickness dependence of the band edges near the gap.

In summary, we have used a recently developed pressure cell measurement setup to investigate the pressure response of ZrTe$_5$ nanodevices in magnetotransport. The multicarrier transport model is suitable for extracting the charge carrier parameters from measurements under pressure. The band fitting analysis using the temperature dependence of carrier density indicates the closing and subsequent reopening of the direct gap at the $\Upgamma$ point with pressure. This is consistent with expectations from a WTI-to-STI phase transition. In addition, a reduction with pressure of a $B_\text{z}$-independent carrier contribution $\sigma_1$ to the conductance is observed. This could be a further indication of the phase transition. Such a contribution could be caused in part by the edge states running along the ZrTe$_5$ layers, while increased backscattering under pressure might decrease $\sigma_1$. In thinner devices, the band fitting analysis is able to show the gradual shift of the chemical potential into the valence band as the thickness is reduced, explaining the lack of electron-dominanted transport in thin ZrTe$_5$ samples at low temperatures.

The experimental results under pressure can be compared favorably with expectations from band structure calculations along lattice compression paths. This joint theoretical and experimental approach suggests the existence of a phase transition driven by hydrostatic pressure from WTI to STI phase in particular. According to the observed evolution of the gap at $\Upgamma$ with pressure, in conjunction with the strain-dependent phase diagram, in case of isotropic compression the phase transition corresponds to only an about 0.2\% change in lattice parameters, while for a compression purely in the vdW stacking direction 0.9\% is required. This is consistent with our finding of a robust resistivity minimum at room temperature, at approximately 0.7 GPa, which has been previously attributed to a topological phase transition related to the gap at the $\Upgamma$ point\,\cite{dirac-and-sti-and-wti_mutch-strainTunedTransition-esttptz_2019}. Such a phase transition would offer further experimental support for the observation that our CVT-grown ZrTe$_5$ is in a WTI phase at ambient pressure, because hydrostatic compression can only be expected to drive the phase transition in one direction.

\section*{Methods}

\subsection*{Density functional theory}

The optimized geometry and electronic properties of the crystal were obtained by the SIESTA implementation of DFT\,\cite{artacho2008siesta,soler2002siesta,garcia2020siesta,fernandez2006site}. SIESTA employs norm-conserving pseudopotentials to account for the core electrons and linear combination of atomic orbitals to construct the valence states. The generalised gradient approximation of the exchange and the correlation functional was used with Perdew–Burke–Ernzerhof parametrisation\,\cite{perdew1996generalized} and the pseudopotentials optimized by Rivero \textit{et al}.\,\cite{rivero2015systematic} with a double-$\zeta$ polarised basis set and a realspace grid defined with an equivalent energy cutoff of 350 Ry for the relaxation phase and 900 Ry for the single-point calculations. The Brillouin zone integration was sampled by a 30$\times$30$\times$18 Monkhorst–Pack $k$-grid for both the relaxation and the single-point calculations.\,\cite{monkhorst1976special}. The geometry optimizations were performed until the forces were smaller than 0.1 eV nm$^{-1}$.

\subsection*{Sample fabrication}

Macroscopic ZrTe$_5$ crystals grown by the CVT method were obtained from HQ Graphene. We obtained nanoflakes of ZrTe$_5$ by mechanical exfoliation to a SiO$_2$ substrate, using semiconductor processing tape ELP BT-150P-LC produced by Nitto. The thickness of the nanoflakes was verified using atomic force microscopy. Contacts in Hall bar configuration, with larger current bias contacts along the crystallographic $a$-axis, were formed using electron beam lithography, with a Cr adhesion layer of 10 nm followed by a main Au layer of $80-110$ nm. To prepare a cleaner contact surface between ZrTe$_5$ and the metal, an Ar ion beam milling step was performed immediately before deposition. Large area (approx. $100\times100$\,$\upmu$m) contact pads were prepared on the substrate to facilitate wire bonding.

\subsection*{Transport measurements}

Measurement of the nanodevices under hydrostatic pressure was performed in a piston-cylinder pressure cell with an inner diameter of 6.5\,mm, using Daphne 7373 oil as the pressure medium. The contacts of the devices were wire bonded to a PCB that had been built into the plug of the pressure cell, with electrical feed-through wires going through the plug and out of the sample holder. For further details see Ref.\,\cite{pressure_fulop-vdW-pressure_2021}.

Application of pressure (up to 2 GPa) was performed at room temperature, before inserting the pressure cell into the transport measurement setup. A Cryomagnetics, Inc liquid helium cryostat with superconducting magnet (up to a field of 8\,T in the out-of-plane orientation) was used for the magnetotransport measurements. The setup featured a variable temperature insert, into which the pressure cell could fit, in order to achieve temperature stabilization from 1.5 to 300\,K. For the current biasing and voltage readout, low frequency AC measurement techniques were employed using SR830 lock-in amplifiers by Stanford Research Systems.

\section*{Data Availability}

Extended data of the DFT calculations, and an interactive way to browse them, are available online at \href{http://tajkov.ek-cer.hu/zrte5phasediagram/}{tajkov.ek-cer.hu/zrte5phasediagram/}. Further data that support the findings of this study are available from the corresponding author upon request.

\section*{Acknowledgements}

The authors are thankful to the Institute of Technical Physics and Materials Science of the Centre for Energy Research for providing their facilities for sample fabrication, and Z. Balogh, F. Fülöp, M. Hajdu for their technical support. This work has received funding from the FLAG-ERA MultiSpin network, the ERC Twistrain project, and from OTKA grants No. K-138433, No. K-134437, No. K-131938, No. FK-124723, No. PD-134758, and No. K-142179. We acknowledge COST Action CA 21144 SUPERQUMAP and the support of the National Research, Development and Innovation (NRDI) Office of Hungary and the Hungarian Academy of Sciences through the Bolyai and Bolyai+ scholarships (Grant BO/00242/20/11). This research was supported by the Ministry of Culture and Innovation and the National Research, Development and Innovation Office within the Quantum Information National Laboratory of Hungary (Grant No. 2022-2.1.1-NL-2022-00004). Z.T. acknowledges financial support from Slovak Academy of Sciences project IMPULZ IM-2021-42. S.P.D. further acknowledges funding by 2D TECH VINNOVA center (No. 2019-00068), Swedish Research Council VR project grants (No. 2021–04821) and FLAG-ERA project 2DSOTECH (VR No. 2021-05925). Low T infrastructure was provided by VEKOP-2.3.3-15-2017-00015.

\section*{Competing Interests}
The Authors declare no Competing Financial or Non-Financial Interests.

\section*{Author Contributions}
Z.K.K. and B.K. fabricated the devices, with help from A.M.. Transport measurements were performed by Z.K.K., B.K. and A.M.. Transport data analysis and fitting was performed by Z.K.K., E.T., P.M. and S.C.. DFT calculations were performed by D.N., Z.T., L.O. and J.K.. All authors contributed to the manuscript and discussions. S.C., S.P.D., P.M., P.N.-I. and E.T. planned and guided the project.


\begin{thebibliography}{73}%
\makeatletter
\providecommand \@ifxundefined [1]{%
 \@ifx{#1\undefined}
}%
\providecommand \@ifnum [1]{%
 \ifnum #1\expandafter \@firstoftwo
 \else \expandafter \@secondoftwo
 \fi
}%
\providecommand \@ifx [1]{%
 \ifx #1\expandafter \@firstoftwo
 \else \expandafter \@secondoftwo
 \fi
}%
\providecommand \natexlab [1]{#1}%
\providecommand \enquote  [1]{``#1''}%
\providecommand \bibnamefont  [1]{#1}%
\providecommand \bibfnamefont [1]{#1}%
\providecommand \citenamefont [1]{#1}%
\providecommand \href@noop [0]{\@secondoftwo}%
\providecommand \href [0]{\begingroup \@sanitize@url \@href}%
\providecommand \@href[1]{\@@startlink{#1}\@@href}%
\providecommand \@@href[1]{\endgroup#1\@@endlink}%
\providecommand \@sanitize@url [0]{\catcode `\\12\catcode `\$12\catcode
  `\&12\catcode `\#12\catcode `\^12\catcode `\_12\catcode `\%12\relax}%
\providecommand \@@startlink[1]{}%
\providecommand \@@endlink[0]{}%
\providecommand \url  [0]{\begingroup\@sanitize@url \@url }%
\providecommand \@url [1]{\endgroup\@href {#1}{\urlprefix }}%
\providecommand \urlprefix  [0]{URL }%
\providecommand \Eprint [0]{\href }%
\providecommand \doibase [0]{https://doi.org/}%
\providecommand \selectlanguage [0]{\@gobble}%
\providecommand \bibinfo  [0]{\@secondoftwo}%
\providecommand \bibfield  [0]{\@secondoftwo}%
\providecommand \translation [1]{[#1]}%
\providecommand \BibitemOpen [0]{}%
\providecommand \bibitemStop [0]{}%
\providecommand \bibitemNoStop [0]{.\EOS\space}%
\providecommand \EOS [0]{\spacefactor3000\relax}%
\providecommand \BibitemShut  [1]{\csname bibitem#1\endcsname}%
\let\auto@bib@innerbib\@empty
\bibitem [{\citenamefont {Kane}\ and\ \citenamefont
  {Mele}(2005)}]{Kane_QSHE_2005}%
  \BibitemOpen
  \bibfield  {author} {\bibinfo {author} {\bibfnamefont {C.~L.}\ \bibnamefont
  {Kane}}\ and\ \bibinfo {author} {\bibfnamefont {E.~J.}\ \bibnamefont
  {Mele}},\ }\bibfield  {title} {\bibinfo {title} {{Quantum Spin Hall Effect in
  Graphene}},\ }\href {https://doi.org/10.1103/physrevlett.95.226801}
  {\bibfield  {journal} {\bibinfo  {journal} {Phys. Rev. Lett.}\ }\textbf
  {\bibinfo {volume} {95}},\ \bibinfo {pages} {226801} (\bibinfo {year}
  {2005})}\BibitemShut {NoStop}%
\bibitem [{\citenamefont {Hasan}\ and\ \citenamefont
  {Kane}(2010)}]{topo-colloq}%
  \BibitemOpen
  \bibfield  {author} {\bibinfo {author} {\bibfnamefont {M.~Z.}\ \bibnamefont
  {Hasan}}\ and\ \bibinfo {author} {\bibfnamefont {C.~L.}\ \bibnamefont
  {Kane}},\ }\bibfield  {title} {\bibinfo {title} {{Colloquium: Topological
  insulators}},\ }\href {https://doi.org/10.1103/RevModPhys.82.3045} {\bibfield
   {journal} {\bibinfo  {journal} {Rev. Mod. Phys.}\ }\textbf {\bibinfo
  {volume} {82}},\ \bibinfo {pages} {3045} (\bibinfo {year}
  {2010})}\BibitemShut {NoStop}%
\bibitem [{\citenamefont {Sarma}\ \emph {et~al.}(2015)\citenamefont {Sarma},
  \citenamefont {Freedman},\ and\ \citenamefont {Nayak}}]{Sarma_2015_majorana}%
  \BibitemOpen
  \bibfield  {author} {\bibinfo {author} {\bibfnamefont {S.~D.}\ \bibnamefont
  {Sarma}}, \bibinfo {author} {\bibfnamefont {M.}~\bibnamefont {Freedman}},\
  and\ \bibinfo {author} {\bibfnamefont {C.}~\bibnamefont {Nayak}},\ }\bibfield
   {title} {\bibinfo {title} {Majorana zero modes and topological quantum
  computation},\ }\href {http://dx.doi.org/10.1038/npjqi.2015.1} {\bibfield
  {journal} {\bibinfo  {journal} {npj Quantum Inf.}\ }\textbf {\bibinfo
  {volume} {1}},\ \bibinfo {pages} {15001} (\bibinfo {year}
  {2015})}\BibitemShut {NoStop}%
\bibitem [{\citenamefont {Bernevig}\ and\ \citenamefont
  {Zhang}(2006)}]{Bernevig_QSHE_2006}%
  \BibitemOpen
  \bibfield  {author} {\bibinfo {author} {\bibfnamefont {B.~A.}\ \bibnamefont
  {Bernevig}}\ and\ \bibinfo {author} {\bibfnamefont {S.-C.}\ \bibnamefont
  {Zhang}},\ }\bibfield  {title} {\bibinfo {title} {{Quantum Spin Hall
  Effect}},\ }\href {https://doi.org/10.1103/physrevlett.96.106802} {\bibfield
  {journal} {\bibinfo  {journal} {Phys. Rev. Lett.}\ }\textbf {\bibinfo
  {volume} {96}},\ \bibinfo {pages} {106802} (\bibinfo {year}
  {2006})}\BibitemShut {NoStop}%
\bibitem [{\citenamefont {Bernevig}\ \emph {et~al.}(2006)\citenamefont
  {Bernevig}, \citenamefont {Hughes},\ and\ \citenamefont
  {Zhang}}]{Bernevig_qshehgte_2006}%
  \BibitemOpen
  \bibfield  {author} {\bibinfo {author} {\bibfnamefont {B.~A.}\ \bibnamefont
  {Bernevig}}, \bibinfo {author} {\bibfnamefont {T.~L.}\ \bibnamefont
  {Hughes}},\ and\ \bibinfo {author} {\bibfnamefont {S.-C.}\ \bibnamefont
  {Zhang}},\ }\bibfield  {title} {\bibinfo {title} {{Quantum Spin Hall Effect
  and Topological Phase Transition in {HgTe} Quantum Wells}},\ }\href
  {https://doi.org/10.1126/science.1133734} {\bibfield  {journal} {\bibinfo
  {journal} {Science}\ }\textbf {\bibinfo {volume} {314}},\ \bibinfo {pages}
  {1757} (\bibinfo {year} {2006})}\BibitemShut {NoStop}%
\bibitem [{\citenamefont {König}\ \emph {et~al.}(2007)\citenamefont {König}
  \emph {et~al.}}]{Konig_QSHEhgte_2007}%
  \BibitemOpen
  \bibfield  {author} {\bibinfo {author} {\bibfnamefont {M.}~\bibnamefont
  {König}} \emph {et~al.},\ }\bibfield  {title} {\bibinfo {title} {{Quantum
  Spin Hall Insulator State in {HgTe} Quantum Wells}},\ }\href
  {https://doi.org/10.1126/science.1148047} {\bibfield  {journal} {\bibinfo
  {journal} {Science}\ }\textbf {\bibinfo {volume} {318}},\ \bibinfo {pages}
  {766} (\bibinfo {year} {2007})}\BibitemShut {NoStop}%
\bibitem [{\citenamefont {König}\ \emph {et~al.}(2008)\citenamefont {König}
  \emph {et~al.}}]{Konig_QSHE_2008}%
  \BibitemOpen
  \bibfield  {author} {\bibinfo {author} {\bibfnamefont {M.}~\bibnamefont
  {König}} \emph {et~al.},\ }\bibfield  {title} {\bibinfo {title} {{The
  Quantum Spin Hall Effect: Theory and Experiment}},\ }\href
  {https://doi.org/10.1143/jpsj.77.031007} {\bibfield  {journal} {\bibinfo
  {journal} {J. Phys. Soc. Jpn.}\ }\textbf {\bibinfo {volume} {77}},\ \bibinfo
  {pages} {031007} (\bibinfo {year} {2008})}\BibitemShut {NoStop}%
\bibitem [{\citenamefont {Xia}\ \emph {et~al.}(2009)\citenamefont {Xia} \emph
  {et~al.}}]{Xia_bi2te3_3dti_2009}%
  \BibitemOpen
  \bibfield  {author} {\bibinfo {author} {\bibfnamefont {Y.}~\bibnamefont
  {Xia}} \emph {et~al.},\ }\bibfield  {title} {\bibinfo {title} {{Observation
  of a large-gap topological-insulator class with a single Dirac cone on the
  surface}},\ }\href {https://doi.org/10.1038/nphys1274} {\bibfield  {journal}
  {\bibinfo  {journal} {Nat. Phys.}\ }\textbf {\bibinfo {volume} {5}},\
  \bibinfo {pages} {398} (\bibinfo {year} {2009})}\BibitemShut {NoStop}%
\bibitem [{\citenamefont {Zhang}\ \emph {et~al.}(2009)\citenamefont {Zhang}
  \emph {et~al.}}]{Zhang_bi2te3_theory_2009}%
  \BibitemOpen
  \bibfield  {author} {\bibinfo {author} {\bibfnamefont {H.}~\bibnamefont
  {Zhang}} \emph {et~al.},\ }\bibfield  {title} {\bibinfo {title} {{Topological
  insulators in Bi$_2$Se$_3$, Bi$_2$Te$_3$ and Sb$_2$Te$_3$ with a single Dirac
  cone on the surface}},\ }\href {https://doi.org/10.1038/nphys1270} {\bibfield
   {journal} {\bibinfo  {journal} {Nat. Phys.}\ }\textbf {\bibinfo {volume}
  {5}},\ \bibinfo {pages} {438} (\bibinfo {year} {2009})}\BibitemShut {NoStop}%
\bibitem [{\citenamefont {Qi}\ and\ \citenamefont
  {Zhang}(2011)}]{sml_and_supcond-topo-2011}%
  \BibitemOpen
  \bibfield  {author} {\bibinfo {author} {\bibfnamefont {X.-L.}\ \bibnamefont
  {Qi}}\ and\ \bibinfo {author} {\bibfnamefont {S.-C.}\ \bibnamefont {Zhang}},\
  }\bibfield  {title} {\bibinfo {title} {{Topological insulators and
  superconductors}},\ }\href {https://doi.org/10.1103/RevModPhys.83.1057}
  {\bibfield  {journal} {\bibinfo  {journal} {Rev. Mod. Phys.}\ }\textbf
  {\bibinfo {volume} {83}},\ \bibinfo {pages} {1057} (\bibinfo {year}
  {2011})}\BibitemShut {NoStop}%
\bibitem [{\citenamefont {Pesin}\ and\ \citenamefont
  {Macdonald}(2012)}]{sml-topo-2012}%
  \BibitemOpen
  \bibfield  {author} {\bibinfo {author} {\bibfnamefont {D.}~\bibnamefont
  {Pesin}}\ and\ \bibinfo {author} {\bibfnamefont {A.}~\bibnamefont
  {Macdonald}},\ }\bibfield  {title} {\bibinfo {title} {{Spintronics and
  Pseudospintronics in Graphene and Topological Insulators}},\ }\href
  {https://doi.org/10.1038/nmat3305} {\bibfield  {journal} {\bibinfo  {journal}
  {Nat. Mater.}\ }\textbf {\bibinfo {volume} {11}},\ \bibinfo {pages} {409}
  (\bibinfo {year} {2012})}\BibitemShut {NoStop}%
\bibitem [{\citenamefont {Tian}\ \emph {et~al.}(2014)\citenamefont {Tian} \emph
  {et~al.}}]{sml-topo-2014}%
  \BibitemOpen
  \bibfield  {author} {\bibinfo {author} {\bibfnamefont {J.}~\bibnamefont
  {Tian}} \emph {et~al.},\ }\bibfield  {title} {\bibinfo {title} {{Topological
  insulator based spin valve devices: Evidence for spin polarized transport of
  spin-momentum-locked topological surface states}},\ }\href
  {https://doi.org/https://doi.org/10.1016/j.ssc.2014.04.005} {\bibfield
  {journal} {\bibinfo  {journal} {Solid State Commun.}\ }\textbf {\bibinfo
  {volume} {191}},\ \bibinfo {pages} {1} (\bibinfo {year} {2014})}\BibitemShut
  {NoStop}%
\bibitem [{\citenamefont {Vaklinova}\ \emph {et~al.}(2016)\citenamefont
  {Vaklinova}, \citenamefont {Hoyer}, \citenamefont {Burghard},\ and\
  \citenamefont {Kern}}]{Vaklinova_grti_2016}%
  \BibitemOpen
  \bibfield  {author} {\bibinfo {author} {\bibfnamefont {K.}~\bibnamefont
  {Vaklinova}}, \bibinfo {author} {\bibfnamefont {A.}~\bibnamefont {Hoyer}},
  \bibinfo {author} {\bibfnamefont {M.}~\bibnamefont {Burghard}},\ and\
  \bibinfo {author} {\bibfnamefont {K.}~\bibnamefont {Kern}},\ }\bibfield
  {title} {\bibinfo {title} {{Current-Induced Spin Polarization in Topological
  Insulator–Graphene Heterostructures}},\ }\href
  {https://doi.org/10.1021/acs.nanolett.6b00167} {\bibfield  {journal}
  {\bibinfo  {journal} {Nano Lett.}\ }\textbf {\bibinfo {volume} {16}},\
  \bibinfo {pages} {2595–2602} (\bibinfo {year} {2016})}\BibitemShut
  {NoStop}%
\bibitem [{\citenamefont {Cha}\ \emph {et~al.}(2018)\citenamefont {Cha} \emph
  {et~al.}}]{optical_bi2se3_sml}%
  \BibitemOpen
  \bibfield  {author} {\bibinfo {author} {\bibfnamefont {S.}~\bibnamefont
  {Cha}} \emph {et~al.},\ }\bibfield  {title} {\bibinfo {title} {{Generation,
  transport and detection of valley-locked spin photocurrent in
  WSe$_2$–graphene–Bi$_2$Se$_3$ heterostructures}},\ }\href
  {https://doi.org/10.1038/s41565-018-0195-y} {\bibfield  {journal} {\bibinfo
  {journal} {Nat. Nanotechnol.}\ }\textbf {\bibinfo {volume} {13}},\ \bibinfo
  {pages} {910–914} (\bibinfo {year} {2018})}\BibitemShut {NoStop}%
\bibitem [{\citenamefont {Khokhriakov}\ \emph {et~al.}(2020)\citenamefont
  {Khokhriakov}, \citenamefont {Hoque}, \citenamefont {Karpiak},\ and\
  \citenamefont {Dash}}]{khokhri-ree-2019}%
  \BibitemOpen
  \bibfield  {author} {\bibinfo {author} {\bibfnamefont {D.}~\bibnamefont
  {Khokhriakov}}, \bibinfo {author} {\bibfnamefont {A.~M.}\ \bibnamefont
  {Hoque}}, \bibinfo {author} {\bibfnamefont {B.}~\bibnamefont {Karpiak}},\
  and\ \bibinfo {author} {\bibfnamefont {S.~P.}\ \bibnamefont {Dash}},\
  }\bibfield  {title} {\bibinfo {title} {{Gate-tunable Spin-Galvanic Effect in
  Graphene Topological insulator van der Waals Heterostructures at Room
  Temperature}},\ }\href {https://doi.org/10.1038/s41467-020-17481-1}
  {\bibfield  {journal} {\bibinfo  {journal} {Nat. Commun.}\ }\textbf {\bibinfo
  {volume} {11}},\ \bibinfo {pages} {3657} (\bibinfo {year}
  {2020})}\BibitemShut {NoStop}%
\bibitem [{\citenamefont {Fjellv{\aa}g}\ and\ \citenamefont
  {Kjekshus}(1986)}]{lattice_fjellvag-powDifExp-spzhpd_1986}%
  \BibitemOpen
  \bibfield  {author} {\bibinfo {author} {\bibfnamefont {H.}~\bibnamefont
  {Fjellv{\aa}g}}\ and\ \bibinfo {author} {\bibfnamefont {A.}~\bibnamefont
  {Kjekshus}},\ }\bibfield  {title} {\bibinfo {title} {{Structural properties
  of ZrTe$_5$ and HfTe$_5$ as seen by powder diffraction}},\ }\href
  {https://doi.org/10.1016/0038-1098(86)90536-3} {\bibfield  {journal}
  {\bibinfo  {journal} {Solid State Commun.}\ }\textbf {\bibinfo {volume}
  {60}},\ \bibinfo {pages} {91} (\bibinfo {year} {1986})}\BibitemShut {NoStop}%
\bibitem [{\citenamefont {DiSalvo}\ \emph {et~al.}(1981)\citenamefont
  {DiSalvo}, \citenamefont {Fleming},\ and\ \citenamefont
  {Waszczak}}]{zrte5_resistanomaly_1981}%
  \BibitemOpen
  \bibfield  {author} {\bibinfo {author} {\bibfnamefont {F.~J.}\ \bibnamefont
  {DiSalvo}}, \bibinfo {author} {\bibfnamefont {R.~M.}\ \bibnamefont
  {Fleming}},\ and\ \bibinfo {author} {\bibfnamefont {J.~V.}\ \bibnamefont
  {Waszczak}},\ }\bibfield  {title} {\bibinfo {title} {{Possible phase
  transition in the quasi-one-dimensional materials ZrTe$_5$ or HfTe$_5$}},\
  }\href {https://doi.org/10.1103/PhysRevB.24.2935} {\bibfield  {journal}
  {\bibinfo  {journal} {Phys. Rev. B}\ }\textbf {\bibinfo {volume} {24}},\
  \bibinfo {pages} {2935} (\bibinfo {year} {1981})}\BibitemShut {NoStop}%
\bibitem [{\citenamefont {Skelton}\ \emph {et~al.}(1982)\citenamefont {Skelton}
  \emph {et~al.}}]{zrte5_resistanomaly_1982}%
  \BibitemOpen
  \bibfield  {author} {\bibinfo {author} {\bibfnamefont {E.}~\bibnamefont
  {Skelton}} \emph {et~al.},\ }\bibfield  {title} {\bibinfo {title} {{Giant
  resistivity and X-ray diffraction anomalies in low-dimensional ZrTe$_5$ and
  HfTe$_5$}},\ }\href
  {https://doi.org/https://doi.org/10.1016/0038-1098(82)91016-X} {\bibfield
  {journal} {\bibinfo  {journal} {Solid State Commun.}\ }\textbf {\bibinfo
  {volume} {42}},\ \bibinfo {pages} {1} (\bibinfo {year} {1982})}\BibitemShut
  {NoStop}%
\bibitem [{\citenamefont {Zhang}\ \emph
  {et~al.}(2017{\natexlab{a}})\citenamefont {Zhang} \emph
  {et~al.}}]{bandstruct-and-fit_zhang-arpeslifshitz-eetilttnz_2017}%
  \BibitemOpen
  \bibfield  {author} {\bibinfo {author} {\bibfnamefont {Y.}~\bibnamefont
  {Zhang}} \emph {et~al.},\ }\bibfield  {title} {{\selectlanguage {english}\bibinfo
  {title} {{Electronic evidence of temperature-induced Lifshitz transition and
  topological nature in ZrTe$_5$}}},\ }\href
  {https://doi.org/10.1038/ncomms15512} {\bibfield  {journal} {\bibinfo
  {journal} {Nat. Commun.}\ }\textbf {\bibinfo {volume} {8}},\ \bibinfo {pages}
  {15512} (\bibinfo {year} {2017}{\natexlab{a}})}\BibitemShut {NoStop}%
\bibitem [{\citenamefont {Wang}\ \emph {et~al.}(2018)\citenamefont {Wang} \emph
  {et~al.}}]{zrte5tp_efshift_2018}%
  \BibitemOpen
  \bibfield  {author} {\bibinfo {author} {\bibfnamefont {W.}~\bibnamefont
  {Wang}} \emph {et~al.},\ }\bibfield  {title} {\bibinfo {title} {{The
  metal-insulator transition in ZrTe$_5$ induced by temperature}},\ }\href
  {https://doi.org/10.1063/1.5064732} {\bibfield  {journal} {\bibinfo
  {journal} {AIP Adv.}\ }\textbf {\bibinfo {volume} {8}},\ \bibinfo {pages}
  {125110} (\bibinfo {year} {2018})}\BibitemShut {NoStop}%
\bibitem [{\citenamefont {Zheng}\ \emph {et~al.}(2016)\citenamefont {Zheng}
  \emph {et~al.}}]{dirac-chiral_guolin-diracMagNeg-te3ddspz_2016}%
  \BibitemOpen
  \bibfield  {author} {\bibinfo {author} {\bibfnamefont {G.}~\bibnamefont
  {Zheng}} \emph {et~al.},\ }\bibfield  {title} {\bibinfo {title} {{Transport
  evidence for the three-dimensional Dirac semimetal phase in ZrTe$_5$}},\
  }\href {https://doi.org/10.1103/PhysRevB.93.115414} {\bibfield  {journal}
  {\bibinfo  {journal} {Phys. Rev. B}\ }\textbf {\bibinfo {volume} {93}},\
  \bibinfo {pages} {115414} (\bibinfo {year} {2016})}\BibitemShut {NoStop}%
\bibitem [{\citenamefont {Shahi}\ \emph {et~al.}(2018)\citenamefont {Shahi}
  \emph {et~al.}}]{fit_shahi-bipolar-bcpoetp_2018}%
  \BibitemOpen
  \bibfield  {author} {\bibinfo {author} {\bibfnamefont {P.}~\bibnamefont
  {Shahi}} \emph {et~al.},\ }\bibfield  {title} {\bibinfo {title} {{Bipolar
  Conduction as the Possible Origin of the Electronic Transition in
  Pentatellurides: Metallic vs Semiconducting Behavior}},\ }\href
  {https://doi.org/10.1103/PhysRevX.8.021055} {\bibfield  {journal} {\bibinfo
  {journal} {Phys. Rev. X}\ }\textbf {\bibinfo {volume} {8}},\ \bibinfo {pages}
  {021055} (\bibinfo {year} {2018})}\BibitemShut {NoStop}%
\bibitem [{\citenamefont {Santos-Cottin}\ \emph {et~al.}(2020)\citenamefont
  {Santos-Cottin} \emph {et~al.}}]{pressure_prb_akrap_2020}%
  \BibitemOpen
  \bibfield  {author} {\bibinfo {author} {\bibfnamefont {D.}~\bibnamefont
  {Santos-Cottin}} \emph {et~al.},\ }\bibfield  {title} {\bibinfo {title}
  {{Probing intraband excitations in ZrTe$_5$: A high-pressure infrared and
  transport study}},\ }\href {https://doi.org/10.1103/PhysRevB.101.125205}
  {\bibfield  {journal} {\bibinfo  {journal} {Phys. Rev. B}\ }\textbf {\bibinfo
  {volume} {101}},\ \bibinfo {pages} {125205} (\bibinfo {year}
  {2020})}\BibitemShut {NoStop}%
\bibitem [{\citenamefont {Morice}\ \emph {et~al.}(2020)\citenamefont {Morice},
  \citenamefont {Lettl}, \citenamefont {Kopp},\ and\ \citenamefont
  {Kampf}}]{zrte5_cvt_fourband}%
  \BibitemOpen
  \bibfield  {author} {\bibinfo {author} {\bibfnamefont {C.}~\bibnamefont
  {Morice}}, \bibinfo {author} {\bibfnamefont {E.}~\bibnamefont {Lettl}},
  \bibinfo {author} {\bibfnamefont {T.}~\bibnamefont {Kopp}},\ and\ \bibinfo
  {author} {\bibfnamefont {A.~P.}\ \bibnamefont {Kampf}},\ }\bibfield  {title}
  {\bibinfo {title} {{Optical conductivity and resistivity in a four-band model
  for ZrTe$_5$ from ab initio calculations}},\ }\href
  {https://link.aps.org/doi/10.1103/PhysRevB.102.155138} {\bibfield  {journal}
  {\bibinfo  {journal} {Phys. Rev. B}\ }\textbf {\bibinfo {volume} {102}},\
  \bibinfo {pages} {155138} (\bibinfo {year} {2020})}\BibitemShut {NoStop}%
\bibitem [{\citenamefont {Weng}\ \emph {et~al.}(2014)\citenamefont {Weng},
  \citenamefont {Dai},\ and\ \citenamefont
  {Fang}}]{bandstruct_weng-QSHparadigm-tmpzh_2014}%
  \BibitemOpen
  \bibfield  {author} {\bibinfo {author} {\bibfnamefont {H.}~\bibnamefont
  {Weng}}, \bibinfo {author} {\bibfnamefont {X.}~\bibnamefont {Dai}},\ and\
  \bibinfo {author} {\bibfnamefont {Z.}~\bibnamefont {Fang}},\ }\bibfield
  {title} {\bibinfo {title} {{Transition-Metal Pentatelluride
  $\mathrm{ZrTe}{}_{5}$ and $\mathrm{HfTe}{}_{5}$: A Paradigm for Large-Gap
  Quantum Spin Hall Insulators}},\ }\href
  {https://doi.org/10.1103/PhysRevX.4.011002} {\bibfield  {journal} {\bibinfo
  {journal} {Phys. Rev. X}\ }\textbf {\bibinfo {volume} {4}},\ \bibinfo {pages}
  {011002} (\bibinfo {year} {2014})}\BibitemShut {NoStop}%
\bibitem [{\citenamefont {Fan}\ \emph {et~al.}(2017)\citenamefont {Fan},
  \citenamefont {Liang}, \citenamefont {Chen}, \citenamefont {Yao},\ and\
  \citenamefont {Zhou}}]{bandstruct_fan-WTIvSTI-tbswtizh_2017}%
  \BibitemOpen
  \bibfield  {author} {\bibinfo {author} {\bibfnamefont {Z.}~\bibnamefont
  {Fan}}, \bibinfo {author} {\bibfnamefont {Q.-F.}\ \bibnamefont {Liang}},
  \bibinfo {author} {\bibfnamefont {Y.~B.}\ \bibnamefont {Chen}}, \bibinfo
  {author} {\bibfnamefont {S.-H.}\ \bibnamefont {Yao}},\ and\ \bibinfo {author}
  {\bibfnamefont {J.}~\bibnamefont {Zhou}},\ }\bibfield  {title}
  {{\selectlanguage {english}\bibinfo {title} {{Transition between strong and weak
  topological insulator in ZrTe$_5$ and HfTe$_5$}}},\ }\href
  {https://doi.org/10.1038/srep45667} {\bibfield  {journal} {\bibinfo
  {journal} {Sci. Rep.}\ }\textbf {\bibinfo {volume} {7}},\ \bibinfo {pages}
  {45667} (\bibinfo {year} {2017})}\BibitemShut {NoStop}%
\bibitem [{\citenamefont {Mutch}\ \emph {et~al.}(2019)\citenamefont {Mutch}
  \emph
  {et~al.}}]{dirac-and-sti-and-wti_mutch-strainTunedTransition-esttptz_2019}%
  \BibitemOpen
  \bibfield  {author} {\bibinfo {author} {\bibfnamefont {J.}~\bibnamefont
  {Mutch}} \emph {et~al.},\ }\bibfield  {title} {\bibinfo {title} {{Evidence
  for a strain-tuned topological phase transition in ZrTe$_5$}},\ }\href
  {https://doi.org/10.1126/sciadv.aav9771} {\bibfield  {journal} {\bibinfo
  {journal} {Sci. Adv.}\ }\textbf {\bibinfo {volume} {5}},\ \bibinfo {pages}
  {eaav9771} (\bibinfo {year} {2019})}\BibitemShut {NoStop}%
\bibitem [{\citenamefont {Wang}\ \emph {et~al.}(2022)\citenamefont {Wang} \emph
  {et~al.}}]{sdho-torus-flux_wang-magchiralaniso-gmatsz_2022}%
  \BibitemOpen
  \bibfield  {author} {\bibinfo {author} {\bibfnamefont {Y.}~\bibnamefont
  {Wang}} \emph {et~al.},\ }\bibfield  {title} {\bibinfo {title} {{Gigantic
  Magnetochiral Anisotropy in the Topological Semimetal
  ${\mathrm{ZrTe}}_{5}$}},\ }\href
  {https://doi.org/10.1103/PhysRevLett.128.176602} {\bibfield  {journal}
  {\bibinfo  {journal} {Phys. Rev. Lett.}\ }\textbf {\bibinfo {volume} {128}},\
  \bibinfo {pages} {176602} (\bibinfo {year} {2022})}\BibitemShut {NoStop}%
\bibitem [{\citenamefont {Manzoni}\ \emph {et~al.}(2016)\citenamefont {Manzoni}
  \emph {et~al.}}]{bandstruct_manzoni-STIarpes-estipz_2016}%
  \BibitemOpen
  \bibfield  {author} {\bibinfo {author} {\bibfnamefont {G.}~\bibnamefont
  {Manzoni}} \emph {et~al.},\ }\bibfield  {title} {\bibinfo {title} {{Evidence
  for a Strong Topological Insulator Phase in ${\mathrm{ZrTe}}_{5}$}},\ }\href
  {https://doi.org/10.1103/PhysRevLett.117.237601} {\bibfield  {journal}
  {\bibinfo  {journal} {Phys. Rev. Lett.}\ }\textbf {\bibinfo {volume} {117}},\
  \bibinfo {pages} {237601} (\bibinfo {year} {2016})}\BibitemShut {NoStop}%
\bibitem [{\citenamefont {Manzoni}\ \emph {et~al.}(2017)\citenamefont {Manzoni}
  \emph {et~al.}}]{arpes_manzoni-ArpesSTI-tdnmbsz_2017}%
  \BibitemOpen
  \bibfield  {author} {\bibinfo {author} {\bibfnamefont {G.}~\bibnamefont
  {Manzoni}} \emph {et~al.},\ }\bibfield  {title} {\bibinfo {title}
  {{Temperature dependent non-monotonic bands shift in ZrTe$_5$}},\ }\href
  {https://doi.org/10.1016/j.elspec.2016.09.006} {\bibfield  {journal}
  {\bibinfo  {journal} {J. Electron Spectrosc. Relat. Phenom.}\ }\textbf
  {\bibinfo {volume} {219}},\ \bibinfo {pages} {9} (\bibinfo {year}
  {2017})}\BibitemShut {NoStop}%
\bibitem [{\citenamefont {Wang}\ \emph {et~al.}(2021)\citenamefont {Wang} \emph
  {et~al.}}]{dirac-sti-and-sdho_wang-magnetoSdHO-mtestipz_2021}%
  \BibitemOpen
  \bibfield  {author} {\bibinfo {author} {\bibfnamefont {J.}~\bibnamefont
  {Wang}} \emph {et~al.},\ }\bibfield  {title} {\bibinfo {title}
  {{Magneto-transport evidence for strong topological insulator phase in
  ZrTe$_5$}},\ }\href {http://dx.doi.org/10.1038/s41467-021-27119-5} {\bibfield
   {journal} {\bibinfo  {journal} {Nat. Commun.}\ }\textbf {\bibinfo {volume}
  {12}},\ \bibinfo {pages} {6758} (\bibinfo {year} {2021})}\BibitemShut
  {NoStop}%
\bibitem [{\citenamefont {Konstantinova}\ \emph {et~al.}(2020)\citenamefont
  {Konstantinova} \emph
  {et~al.}}]{bandstruct-dirac_sun-pumpProbeDirac-pdsz_2020}%
  \BibitemOpen
  \bibfield  {author} {\bibinfo {author} {\bibfnamefont {T.}~\bibnamefont
  {Konstantinova}} \emph {et~al.},\ }\bibfield  {title} {\bibinfo {title}
  {{Photoinduced Dirac semimetal in ZrTe$_5$}},\ }\href
  {https://doi.org/10.1038/s41535-020-00280-8} {\bibfield  {journal} {\bibinfo
  {journal} {npj Quantum Mater.}\ }\textbf {\bibinfo {volume} {5}},\ \bibinfo
  {pages} {80} (\bibinfo {year} {2020})}\BibitemShut {NoStop}%
\bibitem [{\citenamefont {Zhang}\ \emph {et~al.}(2021)\citenamefont {Zhang}
  \emph {et~al.}}]{Zhang_ArpesWTI_2021}%
  \BibitemOpen
  \bibfield  {author} {\bibinfo {author} {\bibfnamefont {P.}~\bibnamefont
  {Zhang}} \emph {et~al.},\ }\bibfield  {title} {\bibinfo {title} {Observation
  and control of the weak topological insulator state in {ZrTe}5},\ }\href
  {https://doi.org/10.1038%2Fs41467-020-20564-8} {\bibfield  {journal}
  {\bibinfo  {journal} {Nat. Commun.}\ }\textbf {\bibinfo {volume} {12}},\
  \bibinfo {pages} {406} (\bibinfo {year} {2021})}\BibitemShut {NoStop}%
\bibitem [{\citenamefont {Li}\ \emph {et~al.}(2016)\citenamefont {Li} \emph
  {et~al.}}]{stmEdge_xbing-stmWTI-eoteds_2016}%
  \BibitemOpen
  \bibfield  {author} {\bibinfo {author} {\bibfnamefont {X.-B.}\ \bibnamefont
  {Li}} \emph {et~al.},\ }\bibfield  {title} {\bibinfo {title} {{Experimental
  Observation of Topological Edge States at the Surface Step Edge of the
  Topological Insulator ${\mathrm{ZrTe}}_{5}$}},\ }\href
  {https://doi.org/10.1103/PhysRevLett.116.176803} {\bibfield  {journal}
  {\bibinfo  {journal} {Phys. Rev. Lett.}\ }\textbf {\bibinfo {volume} {116}},\
  \bibinfo {pages} {176803} (\bibinfo {year} {2016})}\BibitemShut {NoStop}%
\bibitem [{\citenamefont {Wu}\ \emph {et~al.}(2016)\citenamefont {Wu} \emph
  {et~al.}}]{stmEdge_we-stmWTI-etesleg_2016}%
  \BibitemOpen
  \bibfield  {author} {\bibinfo {author} {\bibfnamefont {R.}~\bibnamefont {Wu}}
  \emph {et~al.},\ }\bibfield  {title} {\bibinfo {title} {{Evidence for
  Topological Edge States in a Large Energy Gap near the Step Edges on the
  Surface of ${\mathrm{ZrTe}}_{5}$}},\ }\href
  {https://doi.org/10.1103/PhysRevX.6.021017} {\bibfield  {journal} {\bibinfo
  {journal} {Phys. Rev. X}\ }\textbf {\bibinfo {volume} {6}},\ \bibinfo {pages}
  {021017} (\bibinfo {year} {2016})}\BibitemShut {NoStop}%
\bibitem [{\citenamefont {Liang}\ \emph {et~al.}(2018)\citenamefont {Liang}
  \emph {et~al.}}]{LiangAHEzrte5_2018}%
  \BibitemOpen
  \bibfield  {author} {\bibinfo {author} {\bibfnamefont {T.}~\bibnamefont
  {Liang}} \emph {et~al.},\ }\bibfield  {title} {\bibinfo {title} {{Anomalous
  Hall effect in ZrTe$_5$}},\ }\href
  {https://doi.org/10.1038/s41567-018-0078-z} {\bibfield  {journal} {\bibinfo
  {journal} {Nature Physics}\ }\textbf {\bibinfo {volume} {14}},\ \bibinfo
  {pages} {451–455} (\bibinfo {year} {2018})}\BibitemShut {NoStop}%
\bibitem [{\citenamefont {Sun}\ \emph {et~al.}(2020)\citenamefont {Sun} \emph
  {et~al.}}]{bandstruct-and-sdho-pressure_sun-AHEzeeman-lzsiahez_2020}%
  \BibitemOpen
  \bibfield  {author} {\bibinfo {author} {\bibfnamefont {Z.}~\bibnamefont
  {Sun}} \emph {et~al.},\ }\bibfield  {title} {\bibinfo {title} {{Large Zeeman
  splitting induced anomalous Hall effect in ZrTe$_5$}},\ }\href
  {https://doi.org/10.1038/s41535-020-0239-z} {\bibfield  {journal} {\bibinfo
  {journal} {npj Quantum Mater.}\ }\textbf {\bibinfo {volume} {5}},\ \bibinfo
  {pages} {36} (\bibinfo {year} {2020})}\BibitemShut {NoStop}%
\bibitem [{\citenamefont {Gourgout}\ \emph {et~al.}(2022)\citenamefont
  {Gourgout} \emph {et~al.}}]{gourgout_magneticAHEzrte5_2022}%
  \BibitemOpen
  \bibfield  {author} {\bibinfo {author} {\bibfnamefont {A.}~\bibnamefont
  {Gourgout}} \emph {et~al.},\ }\bibfield  {title} {\bibinfo {title} {{Magnetic
  freeze-out and anomalous Hall effect in ZrTe$_5$}},\ }\href
  {https://doi.org/10.1038/s41535-022-00478-y} {\bibfield  {journal} {\bibinfo
  {journal} {npj Quantum Mater.}\ }\textbf {\bibinfo {volume} {7}},\ \bibinfo
  {pages} {71} (\bibinfo {year} {2022})}\BibitemShut {NoStop}%
\bibitem [{\citenamefont {Liu}\ \emph {et~al.}(2023)\citenamefont {Liu} \emph
  {et~al.}}]{zrte5_thingating_2023}%
  \BibitemOpen
  \bibfield  {author} {\bibinfo {author} {\bibfnamefont {Y.}~\bibnamefont
  {Liu}} \emph {et~al.},\ }\bibfield  {title} {\bibinfo {title} {{Gate-Tunable
  Multiband Transport in ZrTe$_5$ Thin Devices}},\ }\href
  {https://doi.org/10.1021/acs.nanolett.3c01528} {\bibfield  {journal}
  {\bibinfo  {journal} {Nano Lett.}\ }\textbf {\bibinfo {volume} {23}},\
  \bibinfo {pages} {5334} (\bibinfo {year} {2023})}\BibitemShut {NoStop}%
\bibitem [{\citenamefont {Chen}\ \emph
  {et~al.}(2015{\natexlab{a}})\citenamefont {Chen} \emph
  {et~al.}}]{dirac_chen-3Ddirac_msllzs3dmdf_2015}%
  \BibitemOpen
  \bibfield  {author} {\bibinfo {author} {\bibfnamefont {R.~Y.}\ \bibnamefont
  {Chen}} \emph {et~al.},\ }\bibfield  {title} {\bibinfo {title}
  {{Magnetoinfrared Spectroscopy of Landau Levels and Zeeman Splitting of
  Three-Dimensional Massless Dirac Fermions in ${\mathrm{ZrTe}}_{5}$}},\ }\href
  {https://doi.org/10.1103/PhysRevLett.115.176404} {\bibfield  {journal}
  {\bibinfo  {journal} {Phys. Rev. Lett.}\ }\textbf {\bibinfo {volume} {115}},\
  \bibinfo {pages} {176404} (\bibinfo {year} {2015}{\natexlab{a}})}\BibitemShut
  {NoStop}%
\bibitem [{\citenamefont {Chen}\ \emph
  {et~al.}(2015{\natexlab{b}})\citenamefont {Chen} \emph
  {et~al.}}]{dirac_chen-3DdiracOptical_oss3ddsz_2015}%
  \BibitemOpen
  \bibfield  {author} {\bibinfo {author} {\bibfnamefont {R.~Y.}\ \bibnamefont
  {Chen}} \emph {et~al.},\ }\bibfield  {title} {\bibinfo {title} {{Optical
  spectroscopy study of the three-dimensional Dirac semimetal
  ${\mathrm{ZrTe}}_{5}$}},\ }\href {https://doi.org/10.1103/PhysRevB.92.075107}
  {\bibfield  {journal} {\bibinfo  {journal} {Phys. Rev. B}\ }\textbf {\bibinfo
  {volume} {92}},\ \bibinfo {pages} {075107} (\bibinfo {year}
  {2015}{\natexlab{b}})}\BibitemShut {NoStop}%
\bibitem [{\citenamefont {Yuan}\ \emph {et~al.}(2016)\citenamefont {Yuan} \emph
  {et~al.}}]{dirac_xiang-diracMagOpt-oq2ddfz_2015}%
  \BibitemOpen
  \bibfield  {author} {\bibinfo {author} {\bibfnamefont {X.}~\bibnamefont
  {Yuan}} \emph {et~al.},\ }\bibfield  {title} {\bibinfo {title} {{Observation
  of quasi-two-dimensional Dirac fermions in ZrTe$_5$}},\ }\href
  {https://doi.org/10.1038/am.2016.166} {\bibfield  {journal} {\bibinfo
  {journal} {NPG Asia Mater.}\ }\textbf {\bibinfo {volume} {8}},\ \bibinfo
  {pages} {e325} (\bibinfo {year} {2016})}\BibitemShut {NoStop}%
\bibitem [{\citenamefont {Chen}\ \emph {et~al.}(2017)\citenamefont {Chen} \emph
  {et~al.}}]{dirac-and-wti_chen-3DDiracIR-sebbi3dmd_2017}%
  \BibitemOpen
  \bibfield  {author} {\bibinfo {author} {\bibfnamefont {Z.-G.}\ \bibnamefont
  {Chen}} \emph {et~al.},\ }\bibfield  {title} {\bibinfo {title}
  {{Spectroscopic evidence for bulk-band inversion and three-dimensional
  massive Dirac fermions in ZrTe$_5$}},\ }\href
  {https://doi.org/10.1073/pnas.1613110114} {\bibfield  {journal} {\bibinfo
  {journal} {Proc. Natl. Acad. Sci}\ }\textbf {\bibinfo {volume} {114}},\
  \bibinfo {pages} {816} (\bibinfo {year} {2017})}\BibitemShut {NoStop}%
\bibitem [{\citenamefont {Tajkov}\ \emph {et~al.}(2022)\citenamefont {Tajkov}
  \emph {et~al.}}]{topo_tajkov-phase-strain_2022}%
  \BibitemOpen
  \bibfield  {author} {\bibinfo {author} {\bibfnamefont {Z.}~\bibnamefont
  {Tajkov}} \emph {et~al.},\ }\bibfield  {title} {\bibinfo {title} {{Revealing
  the topological phase diagram of ZrTe$_5$ using the complex strain fields of
  microbubbles}},\ }\href {https://doi.org/10.1038/s41524-022-00854-z}
  {\bibfield  {journal} {\bibinfo  {journal} {npj Comput. Mater.}\ }\textbf
  {\bibinfo {volume} {8}},\ \bibinfo {pages} {177} (\bibinfo {year}
  {2022})}\BibitemShut {NoStop}%
\bibitem [{\citenamefont {Yankowitz}\ \emph {et~al.}(2018)\citenamefont
  {Yankowitz} \emph {et~al.}}]{Yankowitz_moirePressure_2018}%
  \BibitemOpen
  \bibfield  {author} {\bibinfo {author} {\bibfnamefont {M.}~\bibnamefont
  {Yankowitz}} \emph {et~al.},\ }\bibfield  {title} {\bibinfo {title} {{Dynamic
  band-structure tuning of graphene moir{\'{e}} superlattices with pressure}},\
  }\href {https://doi.org/10.1038/s41586-018-0107-1} {\bibfield  {journal}
  {\bibinfo  {journal} {Nature}\ }\textbf {\bibinfo {volume} {557}},\ \bibinfo
  {pages} {404} (\bibinfo {year} {2018})}\BibitemShut {NoStop}%
\bibitem [{\citenamefont {Fülöp}\ \emph
  {et~al.}(2021{\natexlab{a}})\citenamefont {Fülöp} \emph
  {et~al.}}]{pressure_fulop-vdW-pressure_2021}%
  \BibitemOpen
  \bibfield  {author} {\bibinfo {author} {\bibfnamefont {B.}~\bibnamefont
  {Fülöp}} \emph {et~al.},\ }\bibfield  {title} {\bibinfo {title} {{New
  method of transport measurements on van der Waals heterostructures under
  pressure}},\ }\href {https://doi.org/10.1063/5.0058583} {\bibfield  {journal}
  {\bibinfo  {journal} {J. Appl. Phys.}\ }\textbf {\bibinfo {volume} {130}},\
  \bibinfo {pages} {064303} (\bibinfo {year} {2021}{\natexlab{a}})}\BibitemShut
  {NoStop}%
\bibitem [{\citenamefont {Hromadov\'a}\ \emph {et~al.}(2013)\citenamefont
  {Hromadov\'a}, \citenamefont {Marto\ifmmode~\check{n}\else \v{n}\fi{}\'ak},\
  and\ \citenamefont {Tosatti}}]{pressure_mos2_mi_2013}%
  \BibitemOpen
  \bibfield  {author} {\bibinfo {author} {\bibfnamefont {L.}~\bibnamefont
  {Hromadov\'a}}, \bibinfo {author} {\bibfnamefont {R.}~\bibnamefont
  {Marto\ifmmode~\check{n}\else \v{n}\fi{}\'ak}},\ and\ \bibinfo {author}
  {\bibfnamefont {E.}~\bibnamefont {Tosatti}},\ }\bibfield  {title} {\bibinfo
  {title} {{Structure change, layer sliding, and metallization in high-pressure
  MoS$_2$}},\ }\href {https://doi.org/10.1103/PhysRevB.87.144105} {\bibfield
  {journal} {\bibinfo  {journal} {Phys. Rev. B}\ }\textbf {\bibinfo {volume}
  {87}},\ \bibinfo {pages} {144105} (\bibinfo {year} {2013})}\BibitemShut
  {NoStop}%
\bibitem [{\citenamefont {Munoz}\ \emph {et~al.}(2016)\citenamefont {Munoz},
  \citenamefont {Collado}, \citenamefont {Usaj}, \citenamefont {Sofo},\ and\
  \citenamefont {Balseiro}}]{pressure_bilayer_2016}%
  \BibitemOpen
  \bibfield  {author} {\bibinfo {author} {\bibfnamefont {F.}~\bibnamefont
  {Munoz}}, \bibinfo {author} {\bibfnamefont {H.~P.~O.}\ \bibnamefont
  {Collado}}, \bibinfo {author} {\bibfnamefont {G.}~\bibnamefont {Usaj}},
  \bibinfo {author} {\bibfnamefont {J.~O.}\ \bibnamefont {Sofo}},\ and\
  \bibinfo {author} {\bibfnamefont {C.~A.}\ \bibnamefont {Balseiro}},\
  }\bibfield  {title} {\bibinfo {title} {{Bilayer graphene under pressure:
  Electron-hole symmetry breaking, valley Hall effect, and Landau levels}},\
  }\href {https://doi.org/10.1103/PhysRevB.93.235443} {\bibfield  {journal}
  {\bibinfo  {journal} {Phys. Rev. B}\ }\textbf {\bibinfo {volume} {93}},\
  \bibinfo {pages} {235443} (\bibinfo {year} {2016})}\BibitemShut {NoStop}%
\bibitem [{\citenamefont {Carr}\ \emph {et~al.}(2018)\citenamefont {Carr},
  \citenamefont {Fang}, \citenamefont {Jarillo-Herrero},\ and\ \citenamefont
  {Kaxiras}}]{TBLG_pressure_theory_2018}%
  \BibitemOpen
  \bibfield  {author} {\bibinfo {author} {\bibfnamefont {S.}~\bibnamefont
  {Carr}}, \bibinfo {author} {\bibfnamefont {S.}~\bibnamefont {Fang}}, \bibinfo
  {author} {\bibfnamefont {P.}~\bibnamefont {Jarillo-Herrero}},\ and\ \bibinfo
  {author} {\bibfnamefont {E.}~\bibnamefont {Kaxiras}},\ }\bibfield  {title}
  {\bibinfo {title} {Pressure dependence of the magic twist angle in graphene
  superlattices},\ }\href {https://doi.org/10.1103/PhysRevB.98.085144}
  {\bibfield  {journal} {\bibinfo  {journal} {Phys. Rev. B}\ }\textbf {\bibinfo
  {volume} {98}},\ \bibinfo {pages} {085144} (\bibinfo {year}
  {2018})}\BibitemShut {NoStop}%
\bibitem [{\citenamefont {Tajkov}\ \emph {et~al.}(2019)\citenamefont {Tajkov},
  \citenamefont {Visontai}, \citenamefont {Oroszlány},\ and\ \citenamefont
  {Koltai}}]{tajkov_uniaxial_bitexGraphene_2019}%
  \BibitemOpen
  \bibfield  {author} {\bibinfo {author} {\bibfnamefont {Z.}~\bibnamefont
  {Tajkov}}, \bibinfo {author} {\bibfnamefont {D.}~\bibnamefont {Visontai}},
  \bibinfo {author} {\bibfnamefont {L.}~\bibnamefont {Oroszlány}},\ and\
  \bibinfo {author} {\bibfnamefont {J.}~\bibnamefont {Koltai}},\ }\bibfield
  {title} {\bibinfo {title} {Uniaxial strain induced topological phase
  transition in bismuth–tellurohalide–graphene heterostructures},\ }\href
  {https://doi.org/10.1039/C9NR04519H} {\bibfield  {journal} {\bibinfo
  {journal} {Nanoscale}\ }\textbf {\bibinfo {volume} {11}},\ \bibinfo {pages}
  {12704} (\bibinfo {year} {2019})}\BibitemShut {NoStop}%
\bibitem [{\citenamefont {Yankowitz}\ \emph {et~al.}(2019)\citenamefont
  {Yankowitz} \emph {et~al.}}]{Yankowitz_tblgPressure_2019}%
  \BibitemOpen
  \bibfield  {author} {\bibinfo {author} {\bibfnamefont {M.}~\bibnamefont
  {Yankowitz}} \emph {et~al.},\ }\bibfield  {title} {\bibinfo {title} {Tuning
  superconductivity in twisted bilayer graphene},\ }\href
  {https://doi.org/10.1126/science.aav1910} {\bibfield  {journal} {\bibinfo
  {journal} {Science}\ }\textbf {\bibinfo {volume} {363}},\ \bibinfo {pages}
  {1059} (\bibinfo {year} {2019})}\BibitemShut {NoStop}%
\bibitem [{\citenamefont {Szentp{\'{e}}teri}\ \emph {et~al.}(2021)\citenamefont
  {Szentp{\'{e}}teri} \emph {et~al.}}]{Szentpeteri_tblg_pressure_2021}%
  \BibitemOpen
  \bibfield  {author} {\bibinfo {author} {\bibfnamefont {B.}~\bibnamefont
  {Szentp{\'{e}}teri}} \emph {et~al.},\ }\bibfield  {title} {\bibinfo {title}
  {{Tailoring the Band Structure of Twisted Double Bilayer Graphene with
  Pressure}},\ }\href {https://doi.org/10.1021/acs.nanolett.1c03066} {\bibfield
   {journal} {\bibinfo  {journal} {Nano Lett.}\ }\textbf {\bibinfo {volume}
  {21}},\ \bibinfo {pages} {8777} (\bibinfo {year} {2021})}\BibitemShut
  {NoStop}%
\bibitem [{\citenamefont {Fülöp}\ \emph
  {et~al.}(2021{\natexlab{b}})\citenamefont {Fülöp} \emph
  {et~al.}}]{balint_proximitySOI_2021}%
  \BibitemOpen
  \bibfield  {author} {\bibinfo {author} {\bibfnamefont {B.}~\bibnamefont
  {Fülöp}} \emph {et~al.},\ }\bibfield  {title} {\bibinfo {title} {{Boosting
  proximity spin{\textendash}orbit coupling in graphene/WSe$_2$
  heterostructures via hydrostatic pressure}},\ }\href
  {https://doi.org/10.1038/s41699-021-00262-9} {\bibfield  {journal} {\bibinfo
  {journal} {npj 2D Mater. Appl.}\ }\textbf {\bibinfo {volume} {5}},\ \bibinfo
  {pages} {82} (\bibinfo {year} {2021}{\natexlab{b}})}\BibitemShut {NoStop}%
\bibitem [{\citenamefont {Kedves}\ \emph {et~al.}(2023)\citenamefont {Kedves}
  \emph {et~al.}}]{kedves2023stabilizing}%
  \BibitemOpen
  \bibfield  {author} {\bibinfo {author} {\bibfnamefont {M.}~\bibnamefont
  {Kedves}} \emph {et~al.},\ }\bibfield  {title} {\bibinfo {title}
  {{Stabilizing the Inverted Phase of a WSe$_2$/BLG/WSe$_2$ Heterostructure via
  Hydrostatic Pressure}},\ }\href
  {https://doi.org/10.1021/acs.nanolett.3c03029} {\bibfield  {journal}
  {\bibinfo  {journal} {Nano Lett.}\ }\textbf {\bibinfo {volume} {23}},\
  \bibinfo {pages} {9508} (\bibinfo {year} {2023})}\BibitemShut {NoStop}%
\bibitem [{\citenamefont {Li}\ \emph {et~al.}(2019)\citenamefont {Li} \emph
  {et~al.}}]{pressure_cri3_afmfm}%
  \BibitemOpen
  \bibfield  {author} {\bibinfo {author} {\bibfnamefont {T.}~\bibnamefont {Li}}
  \emph {et~al.},\ }\bibfield  {title} {\bibinfo {title} {{Pressure-controlled
  interlayer magnetism in atomically thin CrI$_3$}},\ }\href
  {https://doi.org/10.1038/s41563-019-0506-1} {\bibfield  {journal} {\bibinfo
  {journal} {Nat. Mater.}\ }\textbf {\bibinfo {volume} {18}},\ \bibinfo {pages}
  {1303} (\bibinfo {year} {2019})}\BibitemShut {NoStop}%
\bibitem [{\citenamefont {Du}\ \emph {et~al.}(2022)\citenamefont {Du} \emph
  {et~al.}}]{topo-magnet-pressure_yuan-eucd2as2_2022}%
  \BibitemOpen
  \bibfield  {author} {\bibinfo {author} {\bibfnamefont {F.}~\bibnamefont {Du}}
  \emph {et~al.},\ }\bibfield  {title} {\bibinfo {title} {{Consecutive
  topological phase transitions and colossal magnetoresistance in a magnetic
  topological semimetal}},\ }\href {https://doi.org/10.1038/s41535-022-00468-0}
  {\bibfield  {journal} {\bibinfo  {journal} {npj Quantum Mater.}\ }\textbf
  {\bibinfo {volume} {7}},\ \bibinfo {pages} {65} (\bibinfo {year}
  {2022})}\BibitemShut {NoStop}%
\bibitem [{\citenamefont {Bi}\ \emph {et~al.}(2022)\citenamefont {Bi} \emph
  {et~al.}}]{topo-magnet-pressure_vohra-eusn2p2_2022}%
  \BibitemOpen
  \bibfield  {author} {\bibinfo {author} {\bibfnamefont {W.}~\bibnamefont {Bi}}
  \emph {et~al.},\ }\bibfield  {title} {\bibinfo {title} {{Drastic enhancement
  of magnetic critical temperature and amorphization in topological magnet
  EuSn$_2$P$_2$ under pressure}},\ }\href
  {https://doi.org/10.1038/s41535-022-00451-9} {\bibfield  {journal} {\bibinfo
  {journal} {npj Quantum Mater.}\ }\textbf {\bibinfo {volume} {7}},\ \bibinfo
  {pages} {43} (\bibinfo {year} {2022})}\BibitemShut {NoStop}%
\bibitem [{\citenamefont {Zhou}\ \emph {et~al.}(2016)\citenamefont {Zhou} \emph
  {et~al.}}]{pressure_pnas_mao_2016}%
  \BibitemOpen
  \bibfield  {author} {\bibinfo {author} {\bibfnamefont {Y.}~\bibnamefont
  {Zhou}} \emph {et~al.},\ }\bibfield  {title} {\bibinfo {title}
  {{Pressure-induced superconductivity in a three-dimensional topological
  material ZrTe$_5$}},\ }\href
  {https://www.pnas.org/doi/abs/10.1073/pnas.1601262113} {\bibfield  {journal}
  {\bibinfo  {journal} {Proc. Natl. Acad. Sci}\ }\textbf {\bibinfo {volume}
  {113}},\ \bibinfo {pages} {2904} (\bibinfo {year} {2016})}\BibitemShut
  {NoStop}%
\bibitem [{\citenamefont {Zhang}\ \emph
  {et~al.}(2017{\natexlab{b}})\citenamefont {Zhang} \emph
  {et~al.}}]{sdho-nontrivial_zhang_dadss_2017}%
  \BibitemOpen
  \bibfield  {author} {\bibinfo {author} {\bibfnamefont {J.~L.}\ \bibnamefont
  {Zhang}} \emph {et~al.},\ }\bibfield  {title} {\bibinfo {title} {{Disruption
  of the Accidental Dirac Semimetal State in ${\mathrm{ZrTe}}_{5}$ under
  Hydrostatic Pressure}},\ }\href
  {https://doi.org/10.1103/PhysRevLett.118.206601} {\bibfield  {journal}
  {\bibinfo  {journal} {Phys. Rev. Lett.}\ }\textbf {\bibinfo {volume} {118}},\
  \bibinfo {pages} {206601} (\bibinfo {year} {2017}{\natexlab{b}})}\BibitemShut
  {NoStop}%
\bibitem [{\citenamefont {Kov\'acs-Krausz}\ \emph {et~al.}(2023)\citenamefont
  {Kov\'acs-Krausz} \emph {et~al.}}]{zrte5_multicarrier_kkz}%
  \BibitemOpen
  \bibfield  {author} {\bibinfo {author} {\bibfnamefont {Z.}~\bibnamefont
  {Kov\'acs-Krausz}} \emph {et~al.},\ }\bibfield  {title} {\bibinfo {title}
  {{Revealing the band structure of ZrTe$_5$ using multicarrier transport}},\
  }\href {https://doi.org/10.1103/PhysRevB.107.075152} {\bibfield  {journal}
  {\bibinfo  {journal} {Phys. Rev. B}\ }\textbf {\bibinfo {volume} {107}},\
  \bibinfo {pages} {075152} (\bibinfo {year} {2023})}\BibitemShut {NoStop}%
\bibitem [{\citenamefont {Qiu}\ \emph {et~al.}(2016)\citenamefont {Qiu} \emph
  {et~al.}}]{multiband-and-sdho_gang_ooeipahmflz_2016}%
  \BibitemOpen
  \bibfield  {author} {\bibinfo {author} {\bibfnamefont {G.}~\bibnamefont
  {Qiu}} \emph {et~al.},\ }\bibfield  {title} {\bibinfo {title} {{Observation
  of Optical and Electrical In-Plane Anisotropy in High-Mobility Few-Layer
  ZrTe$_5$}},\ }\href {https://doi.org/10.1021/acs.nanolett.6b02629} {\bibfield
   {journal} {\bibinfo  {journal} {Nano Lett.}\ }\textbf {\bibinfo {volume}
  {16}},\ \bibinfo {pages} {7364} (\bibinfo {year} {2016})}\BibitemShut
  {NoStop}%
\bibitem [{\citenamefont {Murata}\ \emph {et~al.}(1997)\citenamefont {Murata},
  \citenamefont {Yoshino}, \citenamefont {Yadav}, \citenamefont {Honda},\ and\
  \citenamefont {Shirakawa}}]{daphne_pressuredep}%
  \BibitemOpen
  \bibfield  {author} {\bibinfo {author} {\bibfnamefont {K.}~\bibnamefont
  {Murata}}, \bibinfo {author} {\bibfnamefont {H.}~\bibnamefont {Yoshino}},
  \bibinfo {author} {\bibfnamefont {H.}~\bibnamefont {Yadav}}, \bibinfo
  {author} {\bibfnamefont {Y.}~\bibnamefont {Honda}},\ and\ \bibinfo {author}
  {\bibfnamefont {N.}~\bibnamefont {Shirakawa}},\ }\bibfield  {title} {\bibinfo
  {title} {Pt resistor thermometry and pressure calibration in a clamped
  pressure cell with the medium, daphne 7373},\ }\href
  {https://doi.org/10.1063/1.1148145} {\bibfield  {journal} {\bibinfo
  {journal} {Rev. Sci. Instrum.}\ }\textbf {\bibinfo {volume} {68}},\ \bibinfo
  {pages} {2490} (\bibinfo {year} {1997})}\BibitemShut {NoStop}%
\bibitem [{\citenamefont {Liu}\ \emph {et~al.}(2016)\citenamefont {Liu} \emph
  {et~al.}}]{bandstruct-and-fit_liu-dynmass-zsdmgdsz_2016}%
  \BibitemOpen
  \bibfield  {author} {\bibinfo {author} {\bibfnamefont {Y.}~\bibnamefont
  {Liu}} \emph {et~al.},\ }\bibfield  {title} {{\selectlanguage {english}\bibinfo
  {title} {{Zeeman splitting and dynamical mass generation in Dirac semimetal
  ZrTe$_5$}}},\ }\href {https://doi.org/10.1038/ncomms12516} {\bibfield
  {journal} {\bibinfo  {journal} {Nat. Commun.}\ }\textbf {\bibinfo {volume}
  {7}},\ \bibinfo {pages} {12516} (\bibinfo {year} {2016})}\BibitemShut
  {NoStop}%
\bibitem [{\citenamefont {Lu}\ \emph {et~al.}(2017)\citenamefont {Lu} \emph
  {et~al.}}]{multiband-and-fit_lu-thicknessDep-tttbtns_2017}%
  \BibitemOpen
  \bibfield  {author} {\bibinfo {author} {\bibfnamefont {J.}~\bibnamefont {Lu}}
  \emph {et~al.},\ }\bibfield  {title} {\bibinfo {title} {{Thickness-tuned
  transition of band topology in ZrTe$_5$ nanosheets}},\ }\href
  {https://doi.org/10.1103/PhysRevB.95.125135} {\bibfield  {journal} {\bibinfo
  {journal} {Phys. Rev. B}\ }\textbf {\bibinfo {volume} {95}},\ \bibinfo
  {pages} {125135} (\bibinfo {year} {2017})}\BibitemShut {NoStop}%
\bibitem [{\citenamefont {Niu}\ \emph {et~al.}(2017)\citenamefont {Niu} \emph
  {et~al.}}]{multiband-nano_jj-etnzs_2017}%
  \BibitemOpen
  \bibfield  {author} {\bibinfo {author} {\bibfnamefont {J.}~\bibnamefont
  {Niu}} \emph {et~al.},\ }\bibfield  {title} {\bibinfo {title} {{Electrical
  transport in nanothick ${\mathrm{ZrTe}}_{5}$ sheets: From three to two
  dimensions}},\ }\href {https://doi.org/10.1103/PhysRevB.95.035420} {\bibfield
   {journal} {\bibinfo  {journal} {Phys. Rev. B}\ }\textbf {\bibinfo {volume}
  {95}},\ \bibinfo {pages} {035420} (\bibinfo {year} {2017})}\BibitemShut
  {NoStop}%
\bibitem [{\citenamefont {Tang}\ \emph {et~al.}(2018)\citenamefont {Tang} \emph
  {et~al.}}]{thicknessDepChn_tang-mctzf_2018}%
  \BibitemOpen
  \bibfield  {author} {\bibinfo {author} {\bibfnamefont {F.}~\bibnamefont
  {Tang}} \emph {et~al.},\ }\bibfield  {title} {\bibinfo {title}
  {{Multi-carrier transport in ZrTe$_5$ film}},\ }\href
  {https://doi.org/10.1088/1674-1056/27/8/087307} {\bibfield  {journal}
  {\bibinfo  {journal} {Chin. Phys. B}\ }\textbf {\bibinfo {volume} {27}},\
  \bibinfo {pages} {087307} (\bibinfo {year} {2018})}\BibitemShut {NoStop}%
\bibitem [{\citenamefont {Artacho}\ \emph {et~al.}(2008)\citenamefont {Artacho}
  \emph {et~al.}}]{artacho2008siesta}%
  \BibitemOpen
  \bibfield  {author} {\bibinfo {author} {\bibfnamefont {E.}~\bibnamefont
  {Artacho}} \emph {et~al.},\ }\bibfield  {title} {\bibinfo {title} {{The
  SIESTA method; developments and applicability}},\ }\href
  {https://doi.org/10.1088/0953-8984/20/6/064208} {\bibfield  {journal}
  {\bibinfo  {journal} {J. Phys. Condens. Matter}\ }\textbf {\bibinfo {volume}
  {20}},\ \bibinfo {pages} {064208} (\bibinfo {year} {2008})}\BibitemShut
  {NoStop}%
\bibitem [{\citenamefont {Soler}\ \emph {et~al.}(2002)\citenamefont {Soler}
  \emph {et~al.}}]{soler2002siesta}%
  \BibitemOpen
  \bibfield  {author} {\bibinfo {author} {\bibfnamefont {J.~M.}\ \bibnamefont
  {Soler}} \emph {et~al.},\ }\bibfield  {title} {\bibinfo {title} {{The SIESTA
  method for ab initio order-N materials simulation}},\ }\href
  {https://doi.org/10.1088/0953-8984/14/11/302} {\bibfield  {journal} {\bibinfo
   {journal} {J. Phys. Condens. Matter}\ }\textbf {\bibinfo {volume} {14}},\
  \bibinfo {pages} {2745} (\bibinfo {year} {2002})}\BibitemShut {NoStop}%
\bibitem [{\citenamefont {Garc{\'\i}a}\ \emph {et~al.}(2020)\citenamefont
  {Garc{\'\i}a} \emph {et~al.}}]{garcia2020siesta}%
  \BibitemOpen
  \bibfield  {author} {\bibinfo {author} {\bibfnamefont {A.}~\bibnamefont
  {Garc{\'\i}a}} \emph {et~al.},\ }\bibfield  {title} {\bibinfo {title}
  {Siesta: Recent developments and applications},\ }\href
  {https://doi.org/10.1063/5.0005077} {\bibfield  {journal} {\bibinfo
  {journal} {J. Chem. Phys.}\ }\textbf {\bibinfo {volume} {152}},\ \bibinfo
  {pages} {204108} (\bibinfo {year} {2020})}\BibitemShut {NoStop}%
\bibitem [{\citenamefont {Fern{\'a}ndez-Seivane}\ \emph
  {et~al.}(2006)\citenamefont {Fern{\'a}ndez-Seivane}, \citenamefont
  {Oliveira}, \citenamefont {Sanvito},\ and\ \citenamefont
  {Ferrer}}]{fernandez2006site}%
  \BibitemOpen
  \bibfield  {author} {\bibinfo {author} {\bibfnamefont {L.}~\bibnamefont
  {Fern{\'a}ndez-Seivane}}, \bibinfo {author} {\bibfnamefont {M.~A.}\
  \bibnamefont {Oliveira}}, \bibinfo {author} {\bibfnamefont {S.}~\bibnamefont
  {Sanvito}},\ and\ \bibinfo {author} {\bibfnamefont {J.}~\bibnamefont
  {Ferrer}},\ }\bibfield  {title} {\bibinfo {title} {{On-site approximation for
  spin--orbit coupling in linear combination of atomic orbitals density
  functional methods}},\ }\href
  {https://doi.org/10.1088/0953-8984/19/48/489001} {\bibfield  {journal}
  {\bibinfo  {journal} {J. Phys. Condens. Matter}\ }\textbf {\bibinfo {volume}
  {18}},\ \bibinfo {pages} {7999} (\bibinfo {year} {2006})}\BibitemShut
  {NoStop}%
\bibitem [{\citenamefont {Perdew}\ \emph {et~al.}(1996)\citenamefont {Perdew},
  \citenamefont {Burke},\ and\ \citenamefont
  {Ernzerhof}}]{perdew1996generalized}%
  \BibitemOpen
  \bibfield  {author} {\bibinfo {author} {\bibfnamefont {J.~P.}\ \bibnamefont
  {Perdew}}, \bibinfo {author} {\bibfnamefont {K.}~\bibnamefont {Burke}},\ and\
  \bibinfo {author} {\bibfnamefont {M.}~\bibnamefont {Ernzerhof}},\ }\bibfield
  {title} {\bibinfo {title} {Generalized gradient approximation made simple},\
  }\href {https://doi.org/10.1103/PhysRevLett.77.3865} {\bibfield  {journal}
  {\bibinfo  {journal} {Phys. Rev. Lett.}\ }\textbf {\bibinfo {volume} {77}},\
  \bibinfo {pages} {3865} (\bibinfo {year} {1996})}\BibitemShut {NoStop}%
\bibitem [{\citenamefont {Rivero}\ \emph {et~al.}(2015)\citenamefont {Rivero}
  \emph {et~al.}}]{rivero2015systematic}%
  \BibitemOpen
  \bibfield  {author} {\bibinfo {author} {\bibfnamefont {P.}~\bibnamefont
  {Rivero}} \emph {et~al.},\ }\bibfield  {title} {\bibinfo {title} {{Systematic
  pseudopotentials from reference eigenvalue sets for DFT calculations}},\
  }\href {https://doi.org/10.1016/j.commatsci.2014.11.026} {\bibfield
  {journal} {\bibinfo  {journal} {Comput. Mater. Sci.}\ }\textbf {\bibinfo
  {volume} {98}},\ \bibinfo {pages} {372} (\bibinfo {year} {2015})}\BibitemShut
  {NoStop}%
\bibitem [{\citenamefont {Monkhorst}\ and\ \citenamefont
  {Pack}(1976)}]{monkhorst1976special}%
  \BibitemOpen
  \bibfield  {author} {\bibinfo {author} {\bibfnamefont {H.~J.}\ \bibnamefont
  {Monkhorst}}\ and\ \bibinfo {author} {\bibfnamefont {J.~D.}\ \bibnamefont
  {Pack}},\ }\bibfield  {title} {\bibinfo {title} {{Special points for
  Brillouin-zone integrations}},\ }\href
  {https://doi.org/10.1103/PhysRevB.13.5188} {\bibfield  {journal} {\bibinfo
  {journal} {Phys. Rev. B}\ }\textbf {\bibinfo {volume} {13}},\ \bibinfo
  {pages} {5188} (\bibinfo {year} {1976})}\BibitemShut {NoStop}%
\end{thebibliography}
%

\end{document}